\documentclass{aa}

\newcommand{\up}{u_{\rm p}}
\newcommand{\vs}{V_{\rm s}}
\newcommand{\vimp}{v_{\rm imp}}
\newcommand{\ec}{E_{\rm c}}
\newcommand{\et}{E_{\rm t}}
\newcommand{\pc}{P_{\rm c}}
\newcommand{\pt}{P_{\rm t}}
\newcommand{\s}{{\cal S}}

\usepackage[nooneline]{subfigure}
\subfiguretopcaptrue
\usepackage{comment}
\usepackage{breqn}
\usepackage{graphicx}
\usepackage[varg]{txfonts}
\usepackage[colorlinks=true, linkcolor=blue, citecolor=blue, filecolor=blue, urlcolor=blue]{hyperref}

\begin{document}

   \title{Formation of planetary atmospheres}
   \subtitle{Analytical estimation of vapor production via planetary impacts}

   \author{Ryushi Miyayama
          \inst{1}
          \and
          Hiroshi Kobayashi\inst{1}
          }

   \institute{Department of Physics, Graduate School of Science, Nagoya University, Furo-cho, Chikusa-ku, Nagoya 464-8602\\
              \email{miyayama.ryushi.w1@s.mail.nagoya-u.ac.jp}\\
             \email{hkobayas@nagoya-u.jp}
             }


\abstract{To investigate impact vaporization for planetary atmosphere formation, we have studied the thermodynamic state generated by the shock wave due to a high-velocity impact, called the shock field. We have carried out iSALE simulations for high-velocity vertical impacts using ANEOS for an equation-of-state (EoS) model. To understand the shock fields obtained from simulations, we have investigated the contribution of the thermal and cold terms in the EoS model on the Hugoniot curves. Although the thermal and cold terms are important for the pressure, the internal energy is mainly determined by the thermal term. We thus assume a simple EoS determined by the thermal term and then analytically derive the shock internal-energy field, which reproduces the results of simulations well. Using the analytical solution of internal energy and the Hugoniot curve, we have derived the shock pressure field analytically as well. The analytical solutions for internal energy and pressure are valid even for impact velocities as low as the sound speed. The solution is good for the vertical direction or within the angles of about 60 degrees. We have applied the solution to impact vaporization for the formation of planetary atmospheres. This gives good estimation of reformation of the planetary atmospheres of Earth sized planet. }

   \keywords{planets and satellites: atmospheres -- planets and satellites: interiors -- equation of state -- hydrodynamics -- shock waves}

   \maketitle
%


\section{Introduction} \label{sec:intro}
Planetary impacts are one of the most common phenomena in planet formation. As planetary bodies grow via collisions, their gravitational accelerations are strong, and impact velocities then exceed the sound speed for planet-sized bodies. Such a hyper-velocity impact generates a shock wave, and this shock wave propagates through the planetary bodies, resulting in irreversible heating and compression. The extreme temperature increase causes melting and vaporization of the bodies and, as a consequence, high-speed impacts have produced planetary oceans and atmospheres \citep[e.g., ][]{Lange_Ahrens_1982Icar,Melosh_Vickery_1989Natur,Tonks_melosh_1993}. The amount of melt and vapor production is derived from the maximum pressure distribution formed by the shock wave, called the shock pressure field \citep[e.g., ][]{1972Moon....4..214A,Stewart_Seifter_2008GeoRL,Kraus_Root_2015NatGe}. In addition, the shock pressure field is important for impact disruption and cratering of planetary bodies \citep{Mizutani_takagi_1990,genda15,suetsugu18}. The collisional outcomes affect planet formation and the collisional evolution of small bodies \citep{kobayashi_tanaka10,kobayashi+10,kobayashi11}. 
Therefore, the shock pressure field is important for understanding phase transition and cratering due to collisions.

Impact vaporization has been investigated via experimental approaches \citep[e.g., ][]{Kadono_sugita_2002GeoRL}, only a limited range of impact velocities have been tested. On the other hand, the shock pressure profiles have been thoroughly tested in theoretical studies over a wide velocity range using analytical methods \citep{1982GS...croft,1989icgp.book.....M,Tonks_melosh_1993} and numerical simulations \citep{1977iecp.symp..639A,Pierazzo_Vickery_1995LPI,1997Icar..127..408P,2000Icar..145..252P,2011Icar..214..724K,Monteux_Arkani_2016,Monteux+2019}. 

Impact simulations show that the shock pressure field is determined as follows. In an impact event, the region of highest pressure forms around the impact site. Such a nearly isobaric region in a target is called the isobaric core. The shock pressure then decreases with distance from the impact point due to the shock propagation. The pressure decrease can be approximated as a power-law function of a radial distance from the core center. \citet{1982GS...croft} implemented a semi-analytical model to express the decreasing shock pressure $P$ as a function of radial distance $R$; 
\begin{equation}\label{eq:previous_pressure}
    P(R)= P_0\left(\frac{R_{\rm ic}}{R} \right)^\alpha,
\end{equation}
where $P_0$ is the pressure in the isobaric core and $R_{\rm ic}$ is its radius. Although the exponent $\alpha$ has been thoroughly investigated using numerical simulations \citep[e.g., ][]{1977iecp.symp..639A,1997Icar..127..408P,2011Icar..214..724K,Monteux_Arkani_2016,Monteux+2019}, the exponent depends on the equation of state (EoS) and has huge uncertainty with ranges from 1 to 3. 

In impact simulations, the exponents $\alpha$ of shock-pressure profiles depend on materials, and the dependence mainly comes from solid EoS models. One of the most sophisticated solid EoS models is M-ANEOS \citep{1972snl..rept..714T,2007MAPS...42.2079M,thompson+2019}, which is written using analytic formulae with 44 parameters. It has been difficult to theoretically understand the material dependence of the shock-pressure profile given this high number of parameters in complicated formulae. 

In this paper, we present an investigation of the shock-pressure field via numerical simulations and analytical methods. We perform simulations for vertical impacts between solid bodies via iSALE with ANEOS. To understand the shock-pressure profile, we also investigate the shock internal-energy profile. We evaluate the contribution of cold and thermal terms in ANEOS to pressure and internal energy. We then analytically derive the shock-pressure profile under the assumption based on the result. Section~\ref{sec:methods} provides a description of the solid EoS model, the setup for numerical simulations, and the Hugoniot curves used to determine the shocked state. In Sect.~\ref{sec:numerical}, we show the results of impact simulations for the shock pressure and internal-energy profiles. We also present the justification for our assumption for the derivation of analytical solutions based on the difference between those values. In Sect.~\ref{sec:analytical}, we show how we derive the analytical solutions for the isobaric and decay regions in shocked profiles. In Sect.~\ref{sec:discussion}, we discuss the dependence of the shock pressure on the direction of the wave propagation in the target. In Sect.~\ref{sec:atmos}, we show how we use our solutions to calculate the amount of vapor in an impact event as a function of impact velocity, and then how we use estimate of vapor to calculate the reformation of the atmosphere removed by a giant impact. We summarize our findings in Sect.~\ref{sec:summary}.

\section{Methods} \label{sec:methods}
\subsection{Solid EoS model: ANEOS}\label{subsec:aneos}
The ANEOS model is widely used in planetary studies \citep[e.g., ][]{1997Icar..127..408P,2007MAPS...42.2079M,2011Icar..214..724K}. For impact simulations, we use the EoS table in the ANEOS model, which is designed by FORTRAN code.

The ANEOS model provides pressure from the Helmholtz free energy, and the free energy is composed of three terms, as follows:
\begin{equation} \label{eq:aneos}
F=\ F_{\rm c}(\rho)\ +\ F_{\rm t}(\rho,T)\ +\ F_{\rm e}(\rho,T)
\end{equation}
The first term on the right-hand side in Eq.~\ref{eq:aneos}, $F_{\rm c}$, the so-called the cold term, depends only on density, and is particular to solid materials and originates from interaction between component particles of the body. The second term $F_{\rm t}$ is the thermal term, and comes from kinetic energy of the component particles. The third term $F_{\rm e}$ is provided to represent the pressure of the ionized electrons generated because of the high-temperature and low-density field; it has a minor effect here because temperatures in typical planetary events are much lower than the ionization temperature \citep{2007MAPS...42.2079M}. 

\subsection{Simulation setup}\label{subsec:setup}
The ANEOS model needs a set of input parameters and we use parameter sets for water ice \citep{2001Sci...294.1326T} and basalt \citep{2005..basalt..pierazzo}. We consider impact simulations between the same materials (ice--ice or basalt--basalt). Planetary impacts are oblique in general. However, we perform numerical simulations of vertical impacts, because we focus on the shock physics determining the shock pressure field. We consider collisions between spherical bodies. However, target bodies are much larger than projectile bodies. Target bodies are thus represented by slabs in simulations. 

The shock pressure field caused in an impact event is calculated with the widely used Lagrangian Eulerian finite-volume shock physics code iSALE-2D \citep{Amsden_etal_1980,Ivanov_etal_1997,2006Icar..180..514W}. We consider the material initial temperature is 150\,K. The projectile radius $R_{\rm p}$ is 500\,m in all simulations. The gravitational force is not included, which is a minor effect for the shock-pressure field. 

The iSALE calculation includes Lagrangian tracer particles, and these particles are placed in each cell to record the thermodynamic state as a function of position and time. We investigate the shock pressure fields using the tracer particles. 

\begin{figure}[htbp]
    \centering
    \includegraphics[width=7cm]{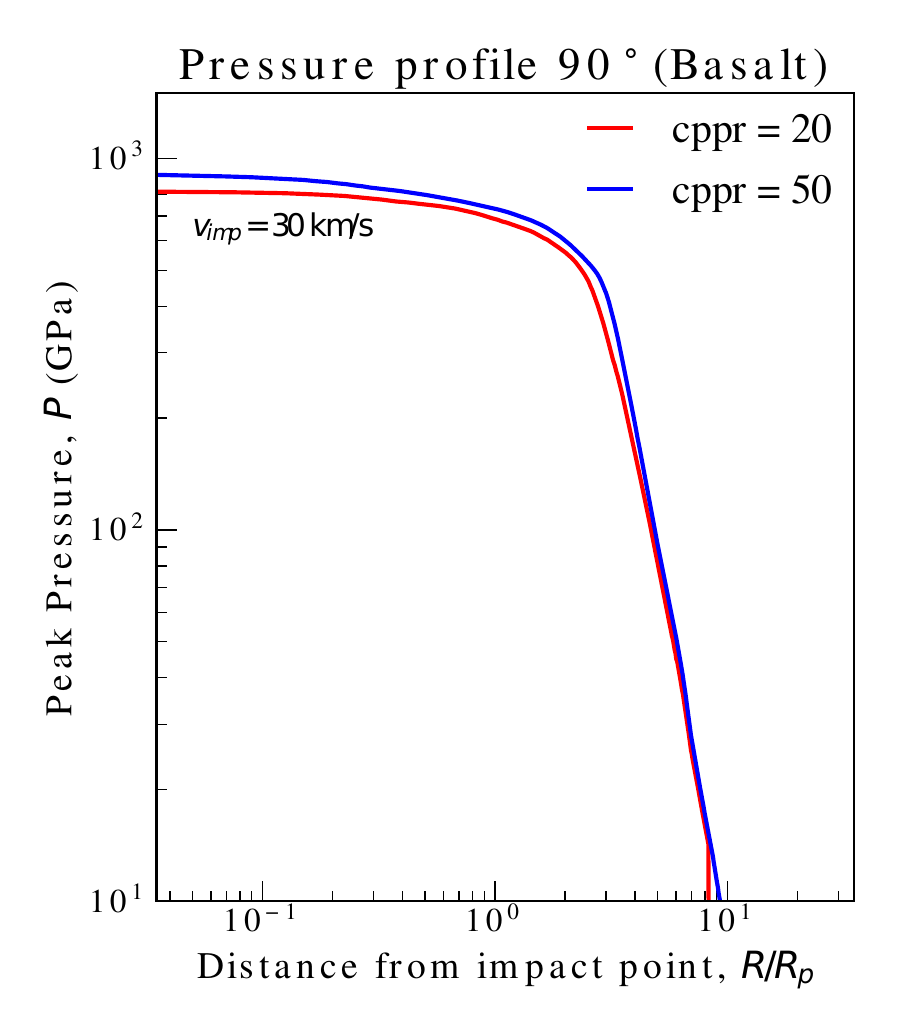}
    \caption{Maximum pressure distribution caused by an impact, also called the shock pressure field, for basalt bodies with an impact speed of 30\,km/s. The red curve is the result for 20 cppr, while the blue curve is for 50 cppr.
}
    \label{fig:cppr_dependence}
\end{figure}

The shock pressure fields obtained from simulations depend on their resolution. In iSALE, the resolution is determined by the cells per projectile radius (cppr). To find a proper resolution, we perform simulations with 20 and 50 cppr (Fig.~\ref{fig:cppr_dependence}). The maximum pressures for 20 cppr are smaller than those for 50 cppr, because low-resolution simulations weaken shock waves numerically. However, the errors for 20 cppr are within 10\% underestimation. To minimize computational time, a fixed resolution of 20 cppr is used in the following simulations.

\section{Numerical results}\label{sec:numerical}
\subsection{Examples of impact simulation}\label{subsec:example}
\begin{figure}[htbp]
\centering
   \subfigure[time: $t/t_{\rm s}=0.0$]{\includegraphics[width=8cm]{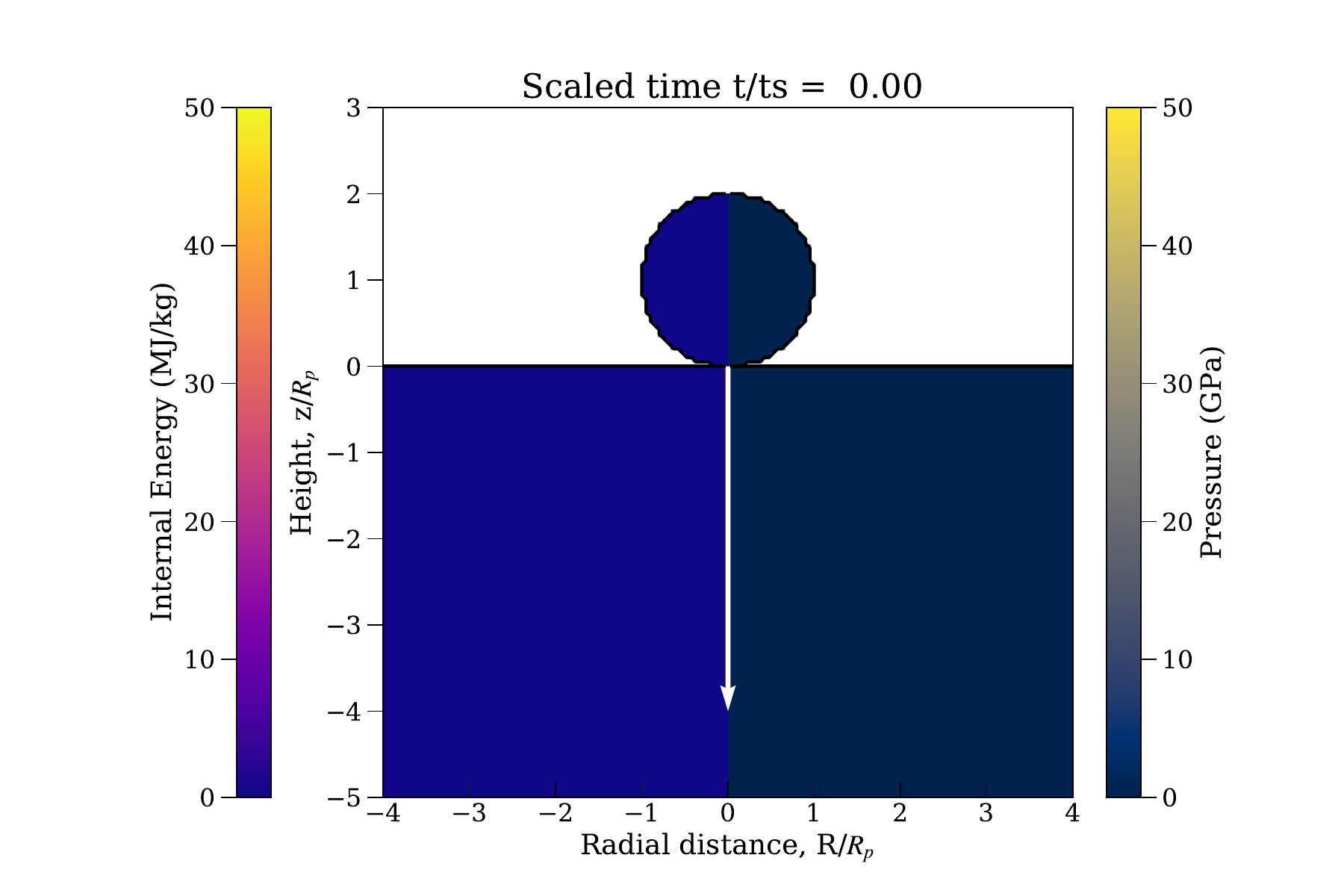}\label{fig:example_t0}}
  \subfigure[time: $t/t_{\rm s}=1.0$]{\includegraphics[width=8cm]{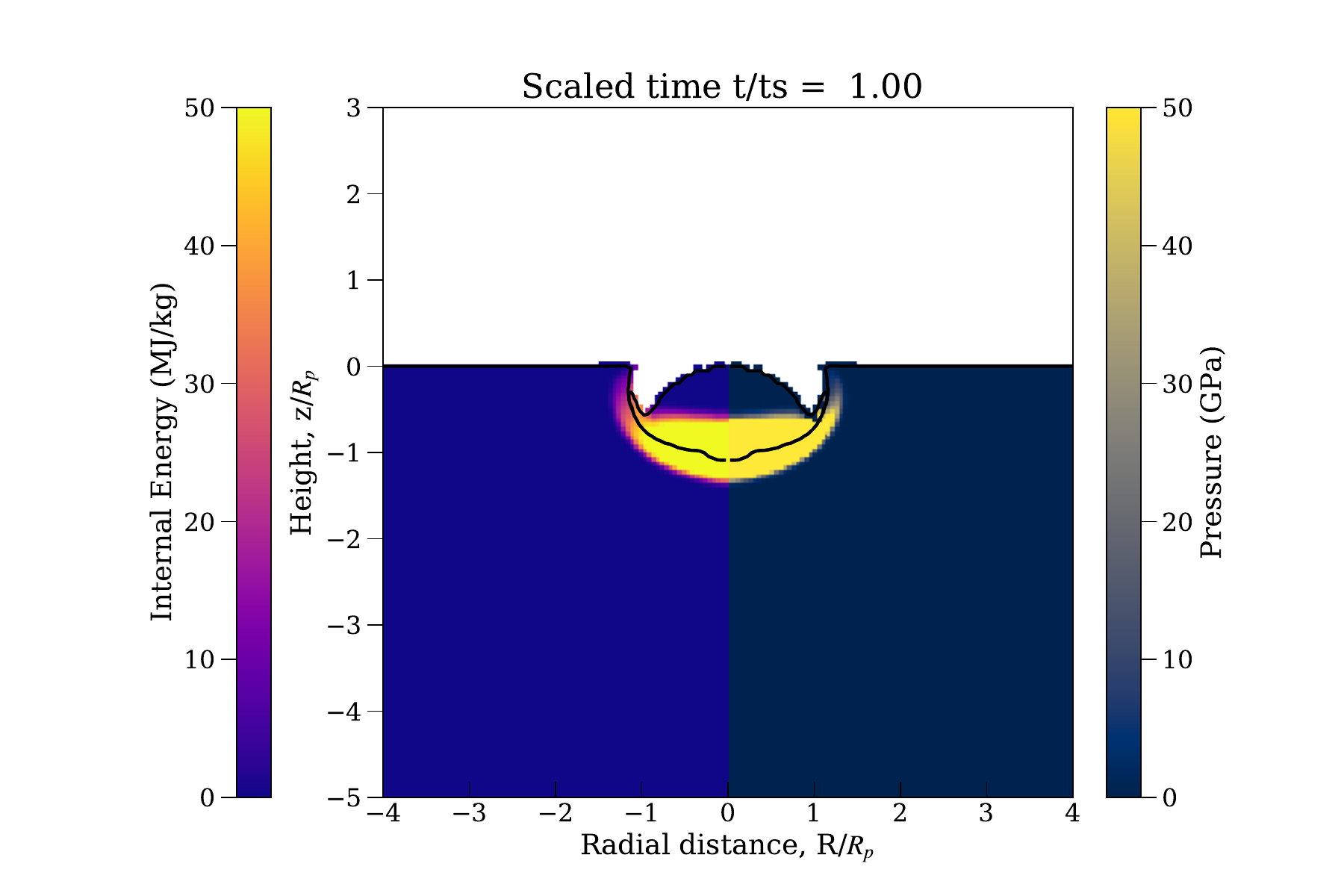}\label{fig:example_t1}}
   \subfigure[time: $t/t_{\rm s}=4.0$]{\includegraphics[width=8cm]{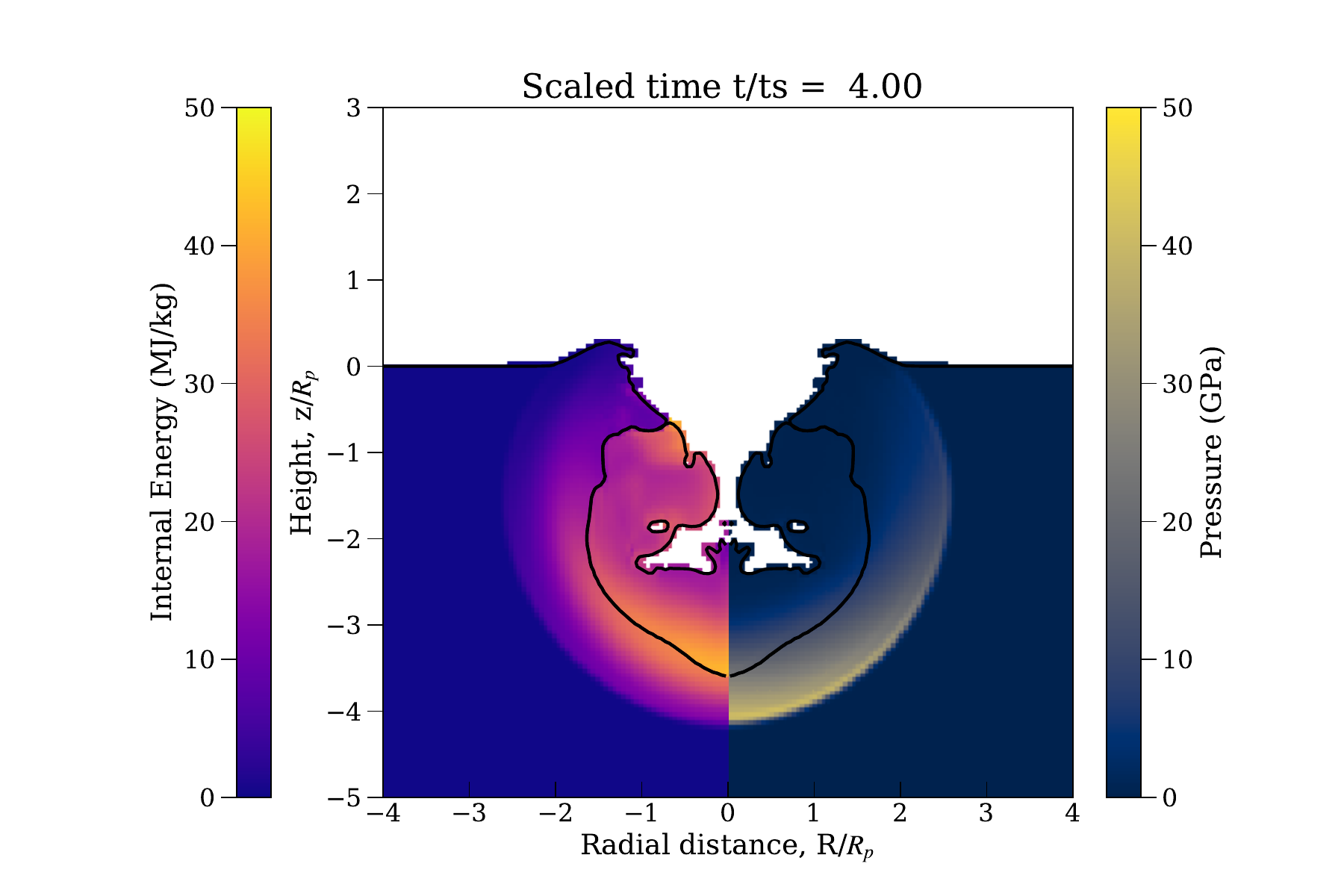}\label{fig:example_t4}}
   \caption{{Result of the iSALE simulation. The changes to pressure and the internal energy caused by an impact shock with $\vimp=30$\,km/s are illustrated as a function of position at $t=0$~(a), $t=t_{\rm s}$ (b), and $t=4t_{\rm s}$ (c).} Calculation time is scaled by the time $t_{\rm s}$, where $t_{\rm s}=2R_{\rm p}/\vimp$. 
\label{fig:example}
}
\end{figure}

Figure~\ref{fig:example} shows the pressure and internal-energy profiles at $t/t_{\rm s}$ = 0, 1, and 4 obtained from an iSALE impact simulation with an impact velocity of $\vimp$ = 30\,km/s between icy bodies, where $t$ is the time from the collision, $t_{\rm s}=2R_{\rm p}/\vimp$, and $R_{\rm p}$ is the radius of the projectile body. The shock wave significantly increases the pressure and internal energy around $|R|\lesssim 2$ and $z\sim -1$ at $t=t_{\rm s}$ (Fig.~\ref{fig:example_t1}), where $R$ is the distance from the symmetric pole and $z$ is the height from the target surface. The shock wave widely propagates and moderately increases the pressure and internal energy at $t=4t_{\rm s}$ (Fig.~\ref{fig:example_t4}), which shows the decay of the shock wave. 

The shock pressure field as shown in Figure 1 is determined by the shock propagation. Pressure reaches a maximum value just after the shock wave has passed. In the early stage, the pressure and internal energy increase around the impact point (Fig.~\ref{fig:example_t1}), while its elevation becomes smaller in a region more distant from the impact point in the later stage (Fig.~\ref{fig:example_t4}). The maximum values of pressure and internal energy are recorded by tracer particles. Figure~\ref{fig:cppr_dependence} is plotted for the shock pressure field along with impact direction, which is shown with a white arrow in Fig.~\ref{fig:example_t0}. In the following sections, we show the vertical shock field for pressure and internal energy. We discuss the direction dependence in Sect.~\ref{sec:discussion}.

\subsection{Shock pressure fields}\label{subsec:result_shockpressure}
Melting and vaporization in an impact event are governed only by the thermodynamic state of the materials formed by shock propagation if the expansion process following shock compression can be assumed to be an adiabatic process. In particular, the shock pressure field has been well investigated because of its importance in the determination of melting, vaporization, and crater size \citep[e.g., ][]{1997Icar..127..408P,2011Icar..214..724K}. 

\begin{figure}[htbp]
\centering
    {\includegraphics[width=7cm]{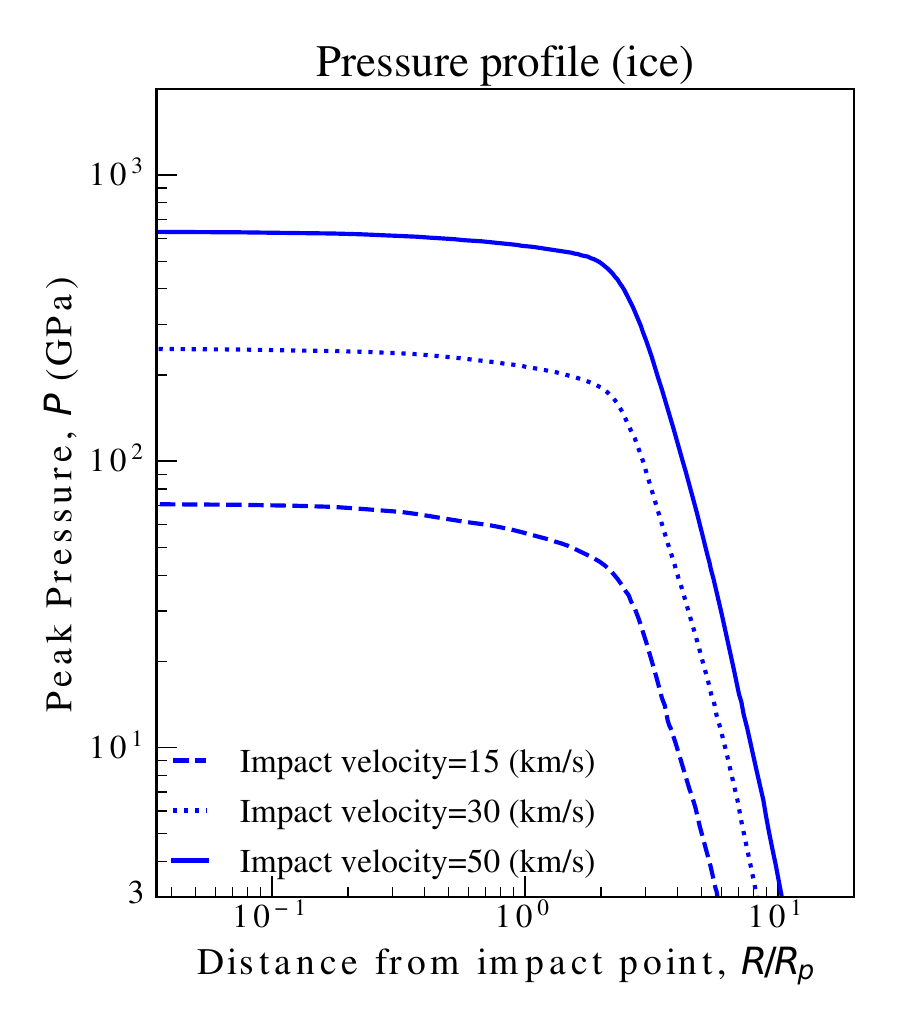}}
    {\includegraphics[width=7cm]{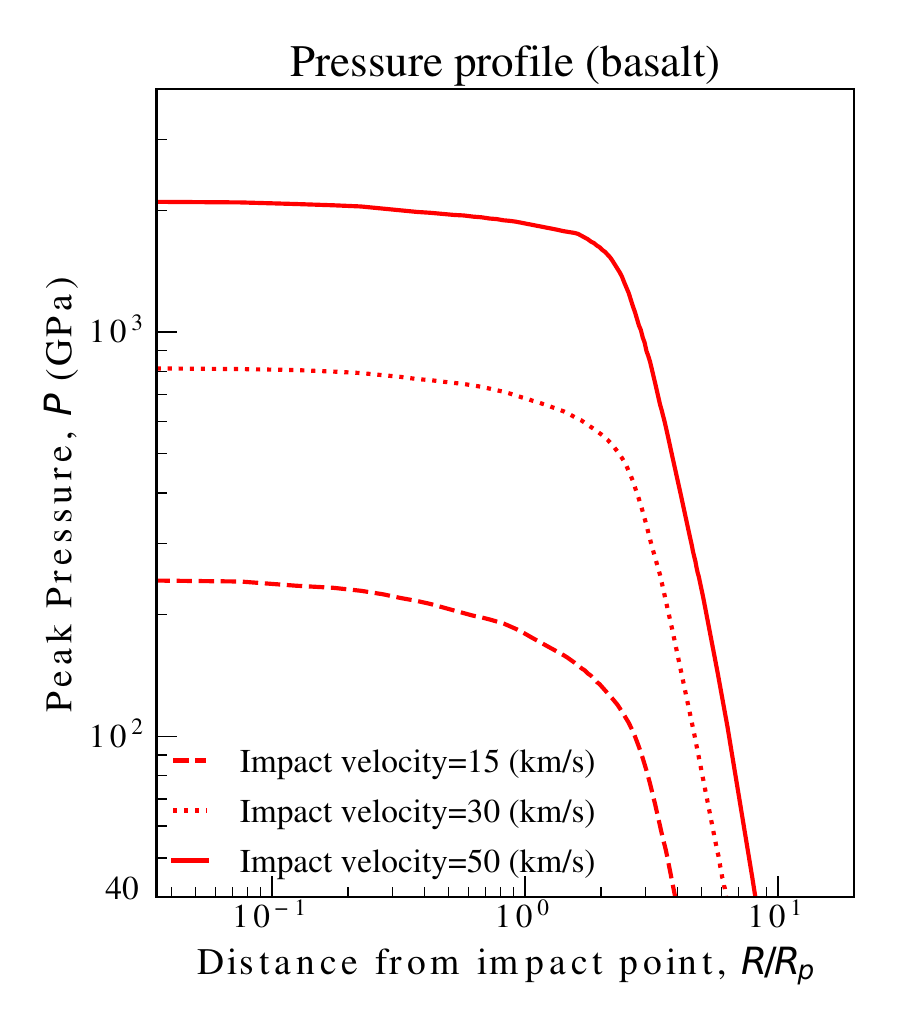}}
   \caption{Peak pressure profile formed by the shock compression as a function of distance from the impact point along the vertical axis. The distance is characterized by projectile radius $R_{\rm p}$. We note that the vertical value, peak pressure, is not a pressure at each time but the maximum pressure reached at each point. Tracer particles along with a white arrow (Fig.~\ref{fig:example}) record maximum pressure as a function of initial position, and these values are shown here. 
\label{fig:prepro_numerical}
}
\end{figure}

Figure~\ref{fig:prepro_numerical} shows the shock pressure fields with distance from the impact point for impacts between icy or basaltic bodies with various impact velocities. The peak pressure is almost the same for $R\lesssim2R_{\rm p}$ around the impact point, which is called the isobaric core. For $R\gg R_{\rm p}$, the peak pressure decreases with distance, which is approximated as a power-law function of $R$.

\subsection{Shock internal energy fields}
\begin{figure}[htbp]
\centering
    {\includegraphics[width=7cm]{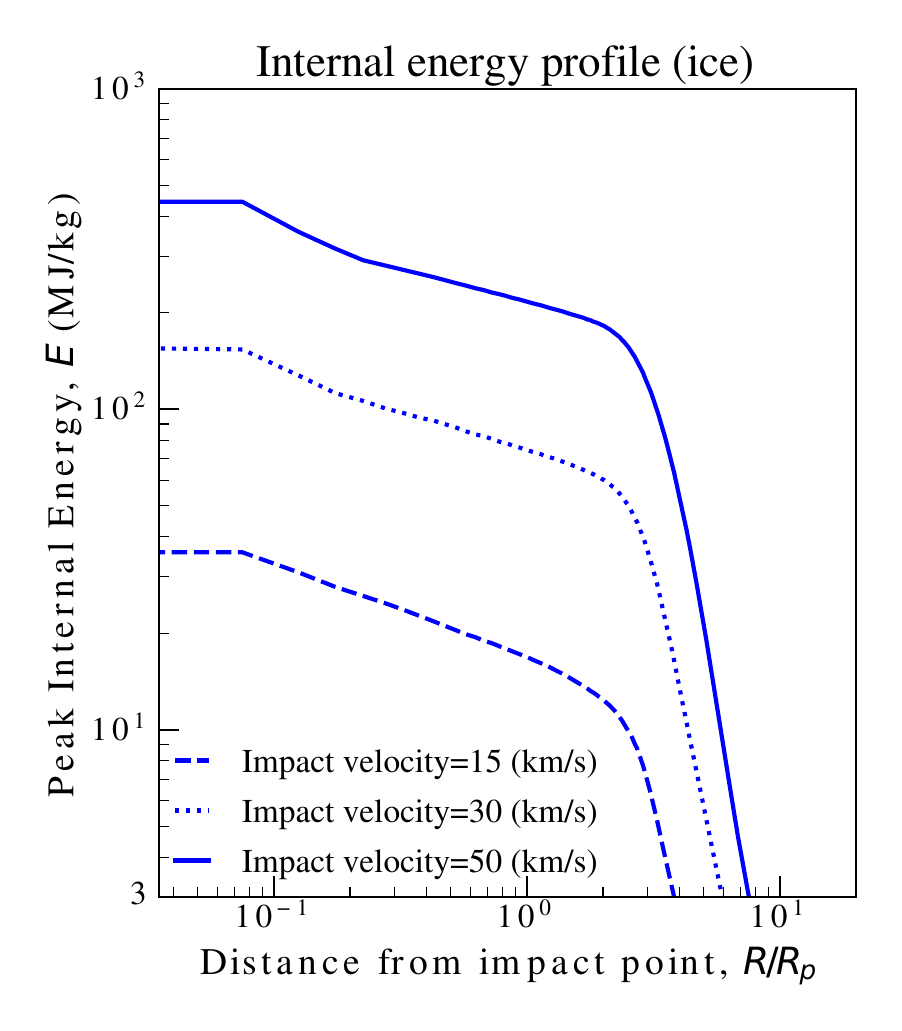}}
    {\includegraphics[width=7cm]{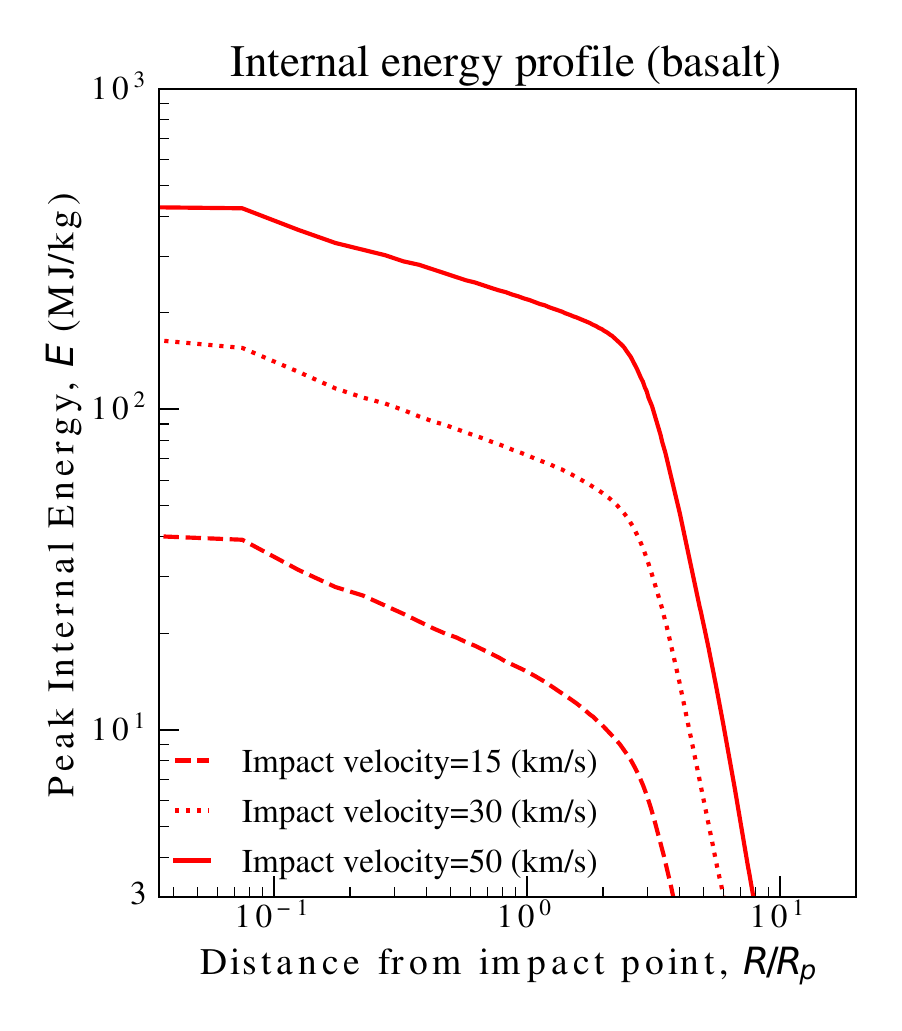}}
   \caption{Same as Fig.~\ref{fig:prepro_numerical}, but for internal energy instead of pressure. 
\label{fig:enepro_numerical_isale}
}
\end{figure}

The shock wave due to an impact event causes an increase in pressure, internal energy, and density. The iSALE package can store the maximum pressure and internal energy of each tracer particle, and provide them as output. The maximum internal-energy profiles and shock internal-energy fields derived from the iSALE simulations are shown in Fig.~\ref{fig:enepro_numerical_isale}. 

The two distributions of pressure and internal energy should have similar profiles as determined by the isobaric core and the radial decay, but the internal energy significantly decreases with distance even in the region corresponding to the isobaric core. This is because the iSALE simulation can provide pressure with high accuracy, but it tends to generate numerical dissipation of internal energy \citep[e.g., ][]{Bowling_Johnson_2020}. 

To evaluate the error on the internal energy, we investigate the shock-tube problem. The states of shocked materials are governed by three conservation laws, called Rankine-Hugoniot equations, and the fundamental equations describing abrupt shock fronts are
\begin{subequations} \label{eq:rh_target}
    \begin{eqnarray}
    \rho(V_{\rm s}-u_{\rm p})&=&\rho_0V_{\rm s},\label{eq:rh-target-mass}\\
    P-P_0&=&\rho_0V_{\rm s}u_{\rm p},\label{eq:rh-target-momentum} \\
    E-E_0&=&\frac{1}{2}(P+P_0)\left(\frac{1}{\rho_0}-\frac{1}{\rho} \right).\label{eq:rh-target-energy}
    \end{eqnarray}
\end{subequations}
These equations involve the pressures $P_0$ and $P$, and the specific internal energies $E_0$ and $E$ in front of and behind the shock wave, respectively, the shock velocity $V_{\rm s}$, the particle velocity $u_{\rm p}$, and the initial density and compressed densities, $\rho_0$ and $\rho$, respectively. Each equation corresponds to a conservation law of mass, momentum, and energy, respectively, across the shock front. A Hugoniot curve is derived from Rankine-Hugoniot equations and the equation of state (e.g., $P=P(\rho,E)$), which represents all of possible states of shocked materials. Therefore, the Hugoniot curve gives the analytical solution of the shock-tube problem.
The Hugoniot curve can be derived from the equations (Eq.~\ref{eq:rh_target} and EoS)), such as $P=P(E),\ V_{\rm s}=V_{\rm s}(u_{\rm p})$. In following sections, we show some results (Figs.~\ref{fig:enepro_numerical} and \ref{fig:hugo_cold}) using this relationship. 

\begin{figure}[htbp]
\centering
   {\includegraphics[width=8cm]{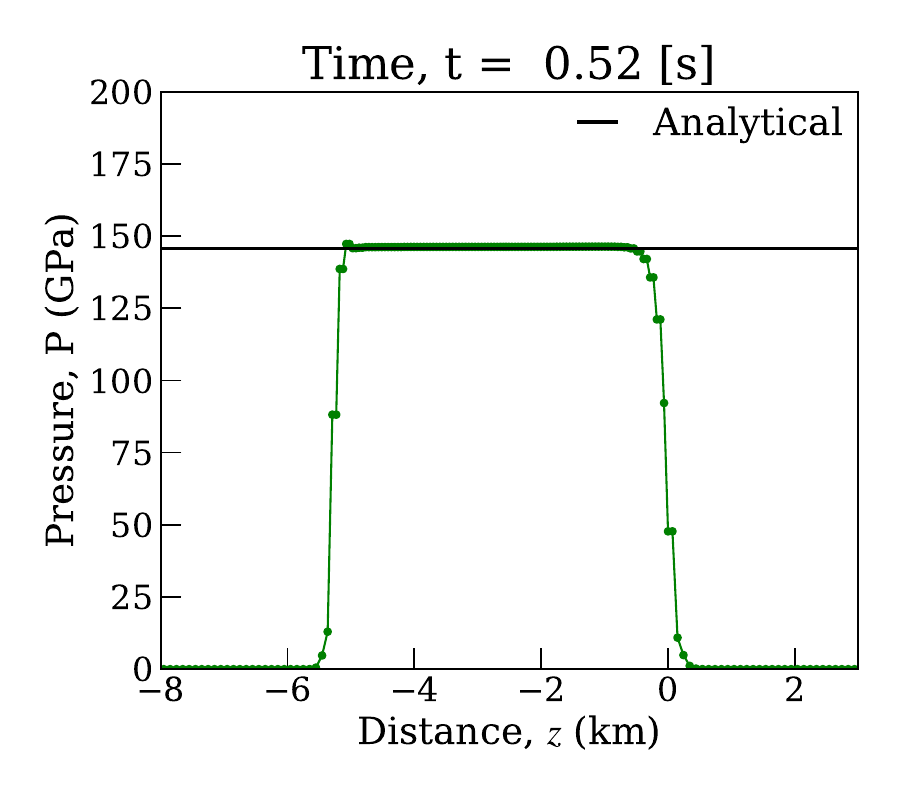}}
  {\includegraphics[width=8cm]{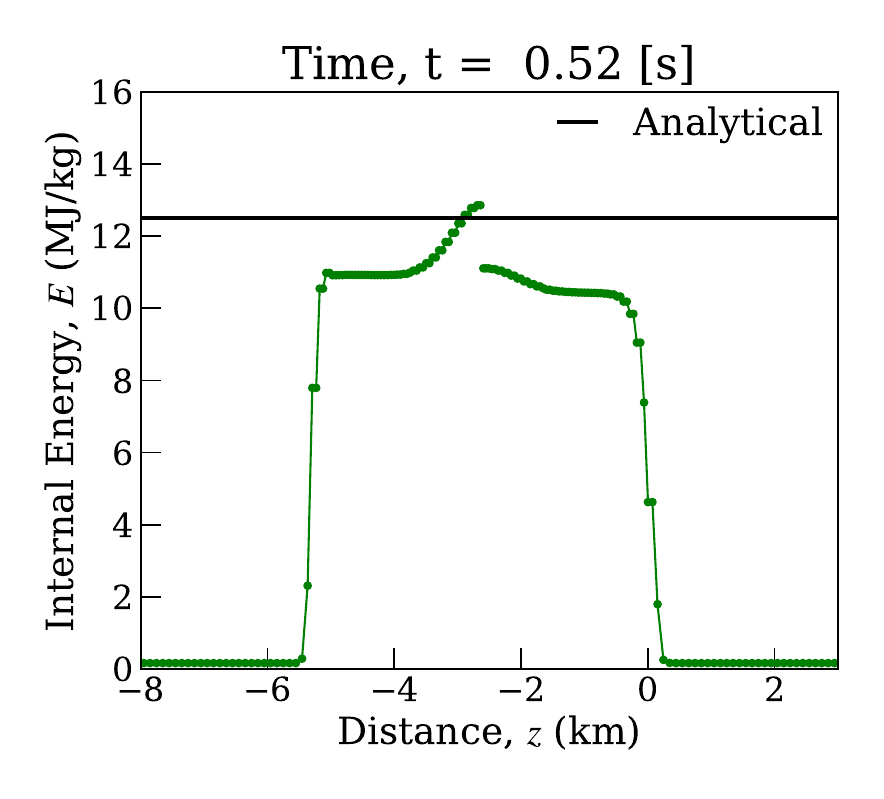}}
\caption{{Snap shots of the pressure and internal energy.} 
Here we show the pressure (a) and internal energy (b) with vertical distance $z$ at time $t$ of 0.52\,s for one-dimensional propagation of a shock wave produced by an impact simulation with a velocity of $\vimp$ = 10\,km/s for a static basaltic cylindrical body at $z\le 0$ vertically colliding with another moving body at $z=0$ at $t=0$. Each green line shows simulation results and black lines correspond to the analytical solutions given by the Hugoniot relation. For $z\approx$ $-6$–0\,km, the pressure and internal energy are increased by the shock wave, which should agree with the analytic solutions.
\label{fig:shocktube}
}
\end{figure}

\begin{figure}[htbp]
\centering
   {\includegraphics[width=7cm]{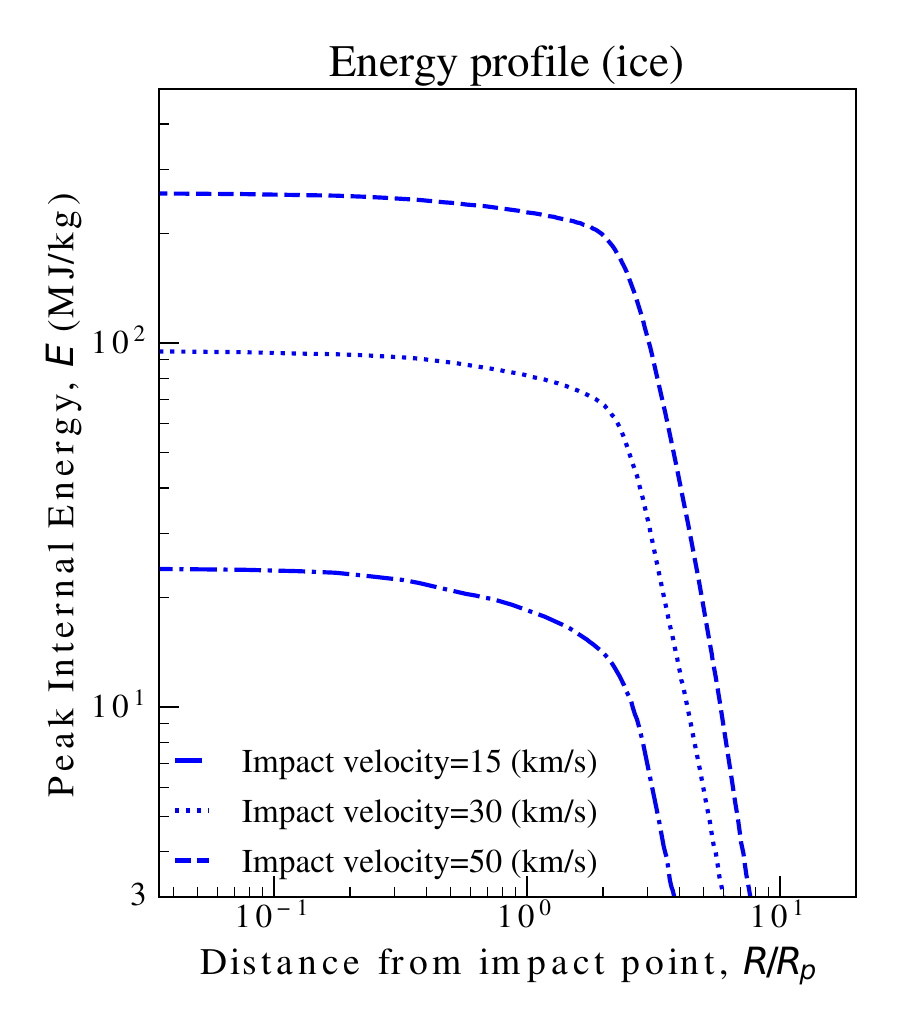}}
   {\includegraphics[width=7cm]{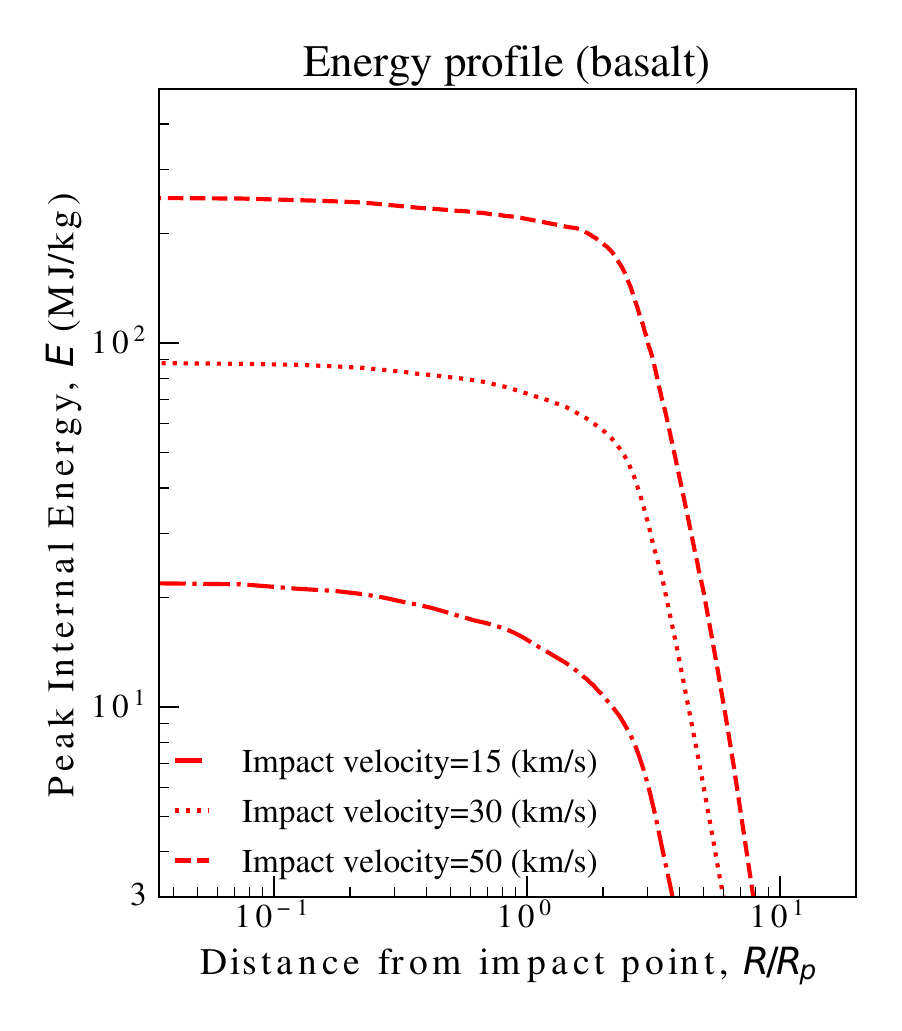}}
\caption{{Peak internal energy profile. We convert shock internal-energy according to the Hugoniot curve ($E=E(P)$). }  
\label{fig:enepro_numerical}
}
\end{figure}

We carried out an iSALE simulation of the shock-tube problem by solving a vertical collision {between basaltic-cylindrical bodies, which have radii of 500\,m. Thus, the collision is between cylinders instead of between a cylinder and a slab.} The numerical and analytical results for pressure and internal energy at each position at a certain time are illustrated in Fig.~\ref{fig:shocktube}. Comparing these results in Fig.~\ref{fig:shocktube}, the iSALE calculation of pressure increasing by the shock wave can represent an analytical solution, but, on the other hand, generated huge errors between numerical and analytical results for internal energy. 

To derive accurate shock internal energy fields with shock pressure fields, we convert shock pressure to shock internal energy using the Hugoniot relation, $E=E(P)$, obtained from Eqs.~\ref{eq:rh-target-mass}-\ref{eq:rh-target-energy} and the EoS. Figure~\ref{fig:enepro_numerical} shows the shock internal-energy field converted using the above method. Thanks to this method, the shock internal energy fields have two structures similar to those in the shock pressure field. In the following sections, we show the internal-energy field obtained using this method.

\subsection{Difference between pressure and internal energy}\label{subsec:heat_capa}
Figure~\ref{fig:enepro_numerical} shows the shock field of the revised internal energy. Its radial profiles are similar to those of the pressure (Fig.~\ref{fig:prepro_numerical}). The values of pressure and internal energy are almost independent of distance for $R\lesssim R_{\rm p}$, while they decrease with $R$ for $R\gtrsim R_{\rm p}$. The radial decay for $R\gtrsim R_{\rm p}$ is approximated as a power-law function of $R$. The power-law exponent at $R\gg R_{\rm p}$ is approximated to $-3$ for the internal energy, while that for the pressure tends to be larger than that for the internal energy. We analytically investigate the shock pressure field in Sect.~\ref{sec:analytical}. 


\section{Analytical expression for the shocked state}\label{sec:analytical}
\subsection{Thermal term versus cold term}\label{subsec:therm_cold}
As mentioned in Sect.~\ref{subsec:aneos}, the ANEOS model provides thermodynamic quantities from the Helmholtz free energy $F$. According to Eq.~(\ref{eq:aneos}), the pressure $P$ and internal energy $E$ are given as: 
\begin{subequations}
\begin{eqnarray}
P\ &=&\ \rho^2\left. \frac{\partial F}{\partial \rho}\right|_T =\ P_{\rm t}(\rho,T)\ +\ \pc(\rho),\label{eq:helm-pre} \\
E\ &=&\ F+TS,\nonumber\\
&=&\ F_{\rm t}(\rho,T)+F_{\rm c}(\rho)+TS(\rho,T),\nonumber\\
&=&\ E_{\rm t}(\rho,T)+E_{\rm c}(\rho),\label{eq:helm-intene}
\end{eqnarray} 
\end{subequations}
where $E_{\rm t}=F_{\rm t}+TS$ and $\ec=F_{\rm c}$, and $S$ is the entropy calculated from $F$ as $S=-(\partial F/\partial T)_\rho$. The pressure and internal energy are composed of the thermal and cold terms, shown in Eqs.~(\ref{eq:helm-pre}) and (\ref{eq:helm-intene}).

\begin{figure*}[htbp]
\centering
   {\includegraphics[width=8cm]{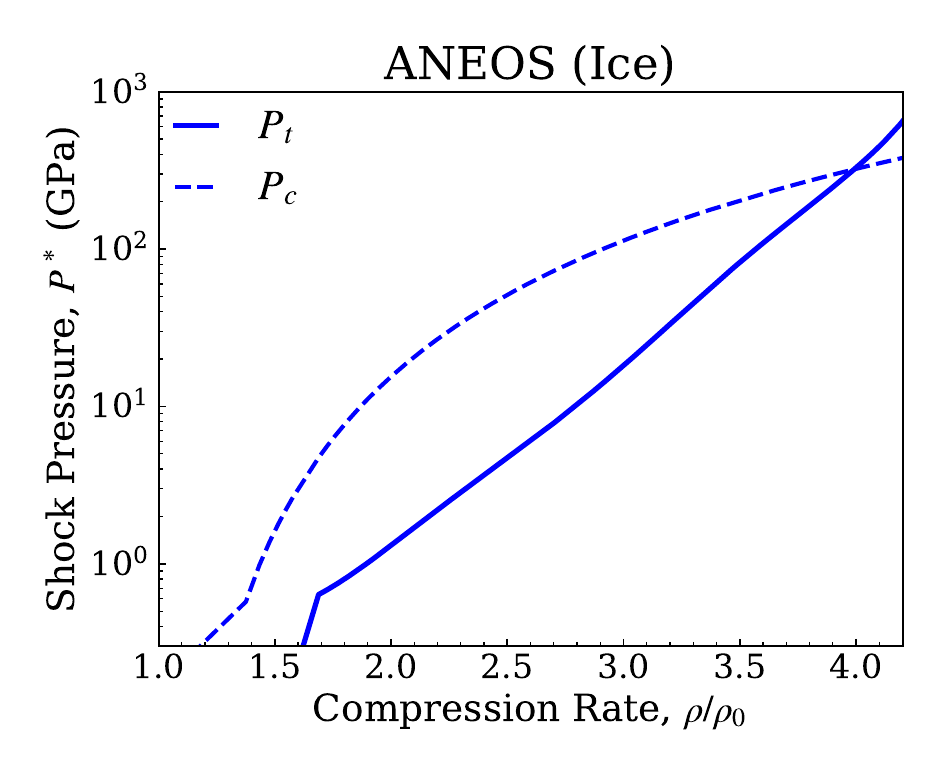}}
   {\includegraphics[width=8cm]{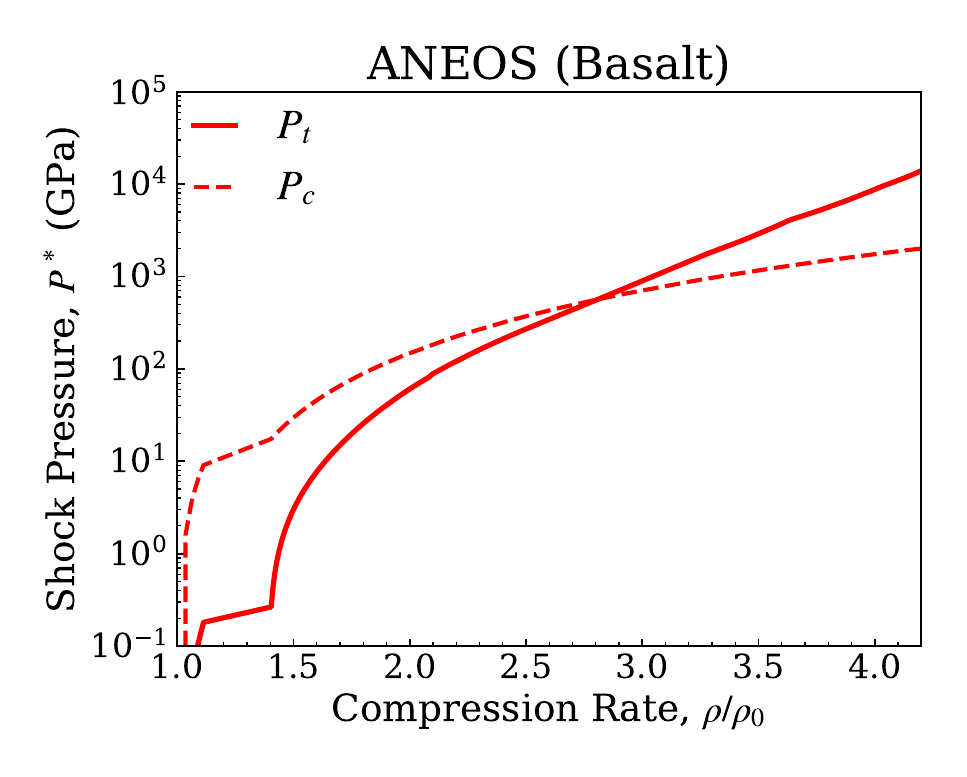}}
   {\includegraphics[width=8cm]{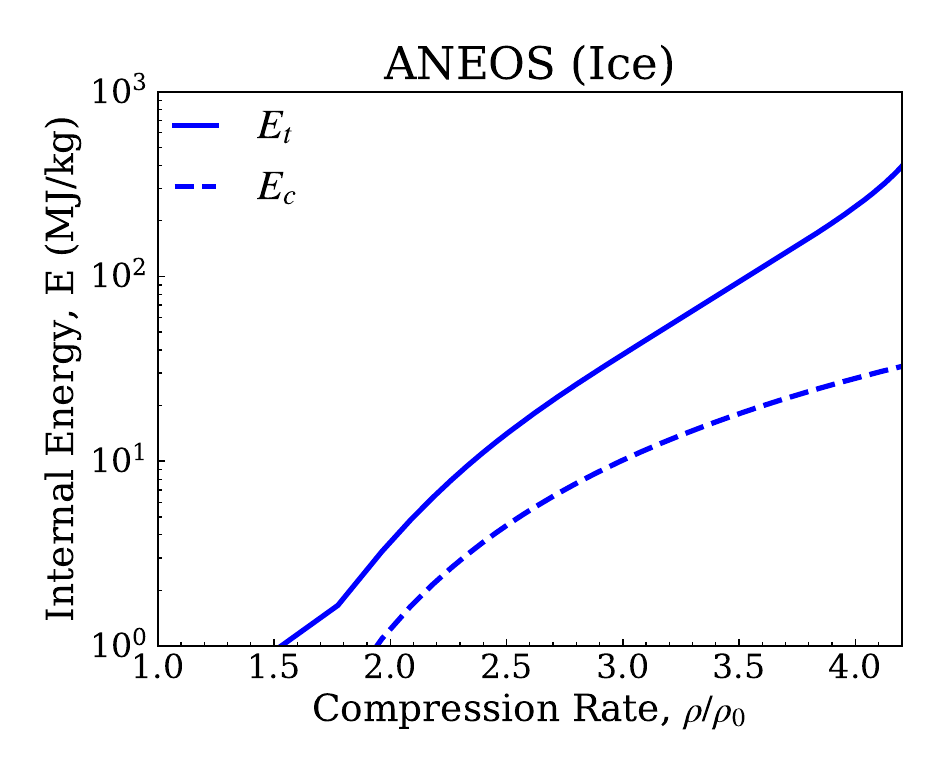}}
   {\includegraphics[width=8cm]{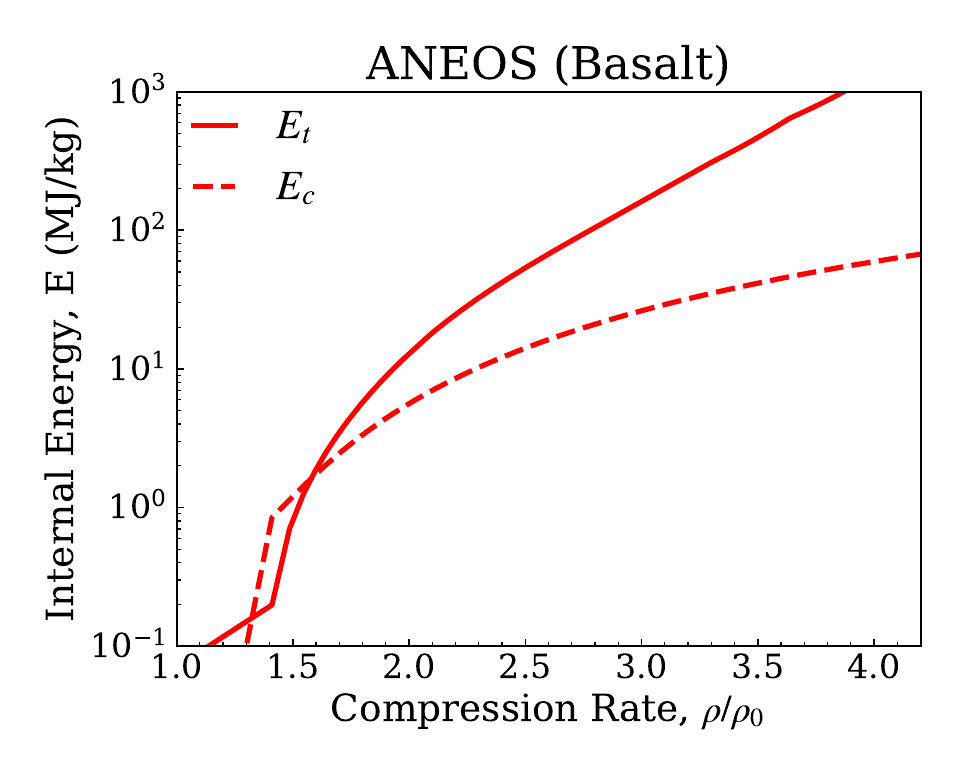}}
  \caption{{Thermal and cold terms in the pressure and the internal energy for ice and basalt.} Each dashed line expresses the cold term and is simply calculated by $P_{\rm c}=P(\rho)$ and $E_{\rm c}=E(\rho)$, and then solid lines are derived by subtracting the cold term from the Hugoniot curve, including the thermal pressure as well. Horizontal values are scaled by the reference density of each material. 
\label{fig:hugo_cold}
}
\end{figure*}

In Fig.~\ref{fig:hugo_cold}, we plot $P_{\rm t}$, $P_{\rm c}$, $\et$, and $\ec$ according to the Hugoniot curves as a function of $\rho$. In Fig.~\ref{fig:hugo_cold}, the cold-term pressure, $\pc$, is dominant for $\rho/\rho_0\lesssim 2\rm -4$, while the thermal-term mainly governs the internal energy even for low density. Therefore, the internal energy is mainly determined by the thermal term.

We focus on the thermal term to derive the shocked internal energy. {According to \citet{2007MAPS...42.2079M},} the definition of $F_{\rm t}$ in the ANEOS model is given by
\begin{eqnarray}
    &&F_{\rm t}(\rho,T) \label{eq:aneos_helm_specific}\\
    \ && = Nk_{\rm B}T\left[3\ln{(1-e^{-T_{\rm D}/T})}-D\left(\frac{T_{\rm D}}{T} \right)+\frac{3}{2}\frac{\ln(1+\psi^b)}{b}\right],\nonumber
\end{eqnarray}    
where $N$ is the total number of atoms per unit mass, $k_{\rm B}$ is the Boltzmann constant, the constant $b$ is a convergence parameter ($0<b\le 1$), $T_{\rm D}$ is the Debye temperature and $D$ is a Debye function, defined as: 
\begin{eqnarray}
    D(x)=\frac{3}{x^3}\int^x_0\frac{y^3}{e^y-1}dy,
\end{eqnarray}    
and $\psi$ is the dimensionless function, difined as: 
\begin{eqnarray}
    \psi(\rho,T)= \frac{C_{13}\rho^{2/3}T}{T_{\rm D}^2}
\end{eqnarray}
where $C_{13}$ is a constant parameter. Based on Eqs.~(\ref{eq:helm-pre}) and (\ref{eq:helm-intene}), the thermal terms for pressure and internal energy are written as:
\begin{subequations}
    \begin{eqnarray}
        P_{\rm t} &=&\rho Nk_{\rm B}T(1+\psi^{-b})^{-1},\label{eq:aneos_pre_specific}\\
        E_{\rm t} &=&3Nk_{\rm B}T\left[D(T_{\rm D}/T)-\frac{1}{2(1+\psi^{-b})}\right].\label{eq:aneos_intene_specific}
    \end{eqnarray}
\end{subequations}
According to Eqs.~(\ref{eq:aneos_pre_specific}) and (\ref{eq:aneos_intene_specific}), $\pt$ and $\et$ are characterized by the Debye temperature, $T_{\rm D}$. 

\begin{figure}[htbp]
    \centering
    \includegraphics[width=8cm]{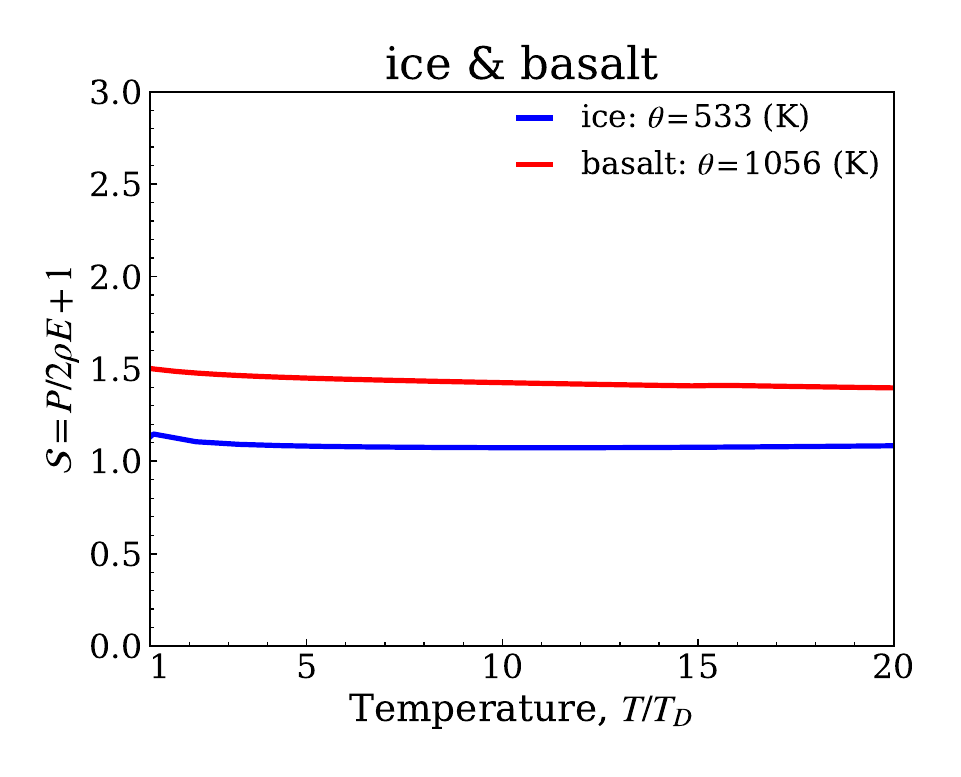}
    \caption{{Dependence of $\s$ on temperature. }    $\s=\pt/2\rho\et+1$ based on the ANEOS Hugoniot curve for water ice (blue) and basalt (red) as a function of the temperature scaled by the Debye temperatures $T_{\rm D}$. }
    \label{fig:heat_cap}
\end{figure}

Behind the strong shock wave with which we are concerned, the temperature is much higher than the Debye temperature. We investigate the dependence of $\s$ on $T$ for $T\gg T_{\rm D}$, where
\begin{eqnarray}
    \s&=& \pt(2\rho \et)^{-1}+1\nonumber\\
    &=& \frac{1}{3(1+\psi^{-b})}[D(T_{\rm D}/T)-(1+\psi^{-b})/2]^{-1}.
\end{eqnarray}
Figure~\ref{fig:heat_cap} shows $\s$ on the Hugoniot curve for water ice and basalt as a function of temperature. As shown in Fig.~\ref{fig:heat_cap}, $\s$ is a constant value dependent on materials.
Therefore, the relation between $\pt$ and $\et$ for $T\gg T_{\rm D}$ is approximately given by
\begin{eqnarray}\label{eq:simple_eos}
    P_{\rm t} \approx 2(\s-1)\rho E_{\rm t}, 
\end{eqnarray}
where $\s=1.26$ and 1.4 for water ice and basalt, respectively. {In Eq.~\ref{eq:simple_eos}, $\s$ can be connected with a Gr\"{u}neisen parameter $\Gamma$ for an ideal gas by $\s=(1+\Gamma)/2$.} In the following section, we show the derivation of the analytical solution for the shock internal energy based on the EoS (Eq.~\ref{eq:simple_eos}).



\subsection{Analytical solution for the shock internal energy}\label{subsec:analytical_ene}
\subsubsection{Isobaric  core}\label{subsubsec:frat}
The particle velocity $\up$ at the isobaric core is obtained from the impedance matching method \citep{1989icgp.book.....M}. For impact between bodies of the same material, the impedance matching gives
\begin{equation}\label{eq:anal_iso_up}
    \up = \vimp/2,
\end{equation}
where $\vimp$ is the impact velocity. Here, we assume $E\approx \et$, $P\approx\pt$, $T\gg T_{\rm D}$ and Eq.~(\ref{eq:simple_eos}) gives the EoS as 
\begin{eqnarray}\label{eq:simple_eos2}
    P= 2(\s-1)\rho E.
\end{eqnarray}
Using Eq.~(\ref{eq:simple_eos2}) and Eqs.~(\ref{eq:rh-target-mass})-(\ref{eq:rh-target-energy}) with $P_0\ll P$ and $E_0\ll E$, we then have
\begin{eqnarray}
    \vs &=& \s \up,\label{eq:anal_iso_shockvelo}\\
    \frac{\rho}{\rho_0}&=&\frac{\s}{\s-1},\label{eq:anal_iso_dens}\\
    P&=&\s\rho_0\up^2,\label{eq:anal_iso_pre}\\
    E&=&\frac{\up^2}{2}\label{eq:anal_iso_ene}.
\end{eqnarray}
We have $P$ and $E$ at the isobaric core, using Eqs.~(\ref{eq:anal_iso_pre}) and (\ref{eq:anal_iso_ene}) coupled with Eq.~(\ref{eq:anal_iso_up}).

The solutions for $P$ and $E$, Eqs.~(\ref{eq:anal_iso_pre}) and (\ref{eq:anal_iso_ene}), are consistent with a one-dimensional shock-tube solution, if high-velocity collisions result in $E\approx\et$ and $P\approx\pt$. In Fig.~\ref{fig:shocktube}, we show the result of the one dimensional problem for basalt with 10\,km/s. Equations~(\ref{eq:anal_iso_pre}), (\ref{eq:anal_iso_ene}), and (\ref{eq:anal_iso_up}) give $P=1.02\times10^2$\,GPa and $E=12.5$\,MJ/kg. The solution for $E$ agrees well with the simulation result, while $P$ given by Eq.~(\ref{eq:anal_iso_pre}) is underestimated. This is mainly caused by the considerable value of $\pc$. Therefore, the assumption of Eq.~(\ref{eq:simple_eos2}) is valid for deriving $E$, while the additional modification is necessary to compensate for the underestimate of $P$.

\begin{figure}[htbp]
    \centering
    \includegraphics[width=7cm]{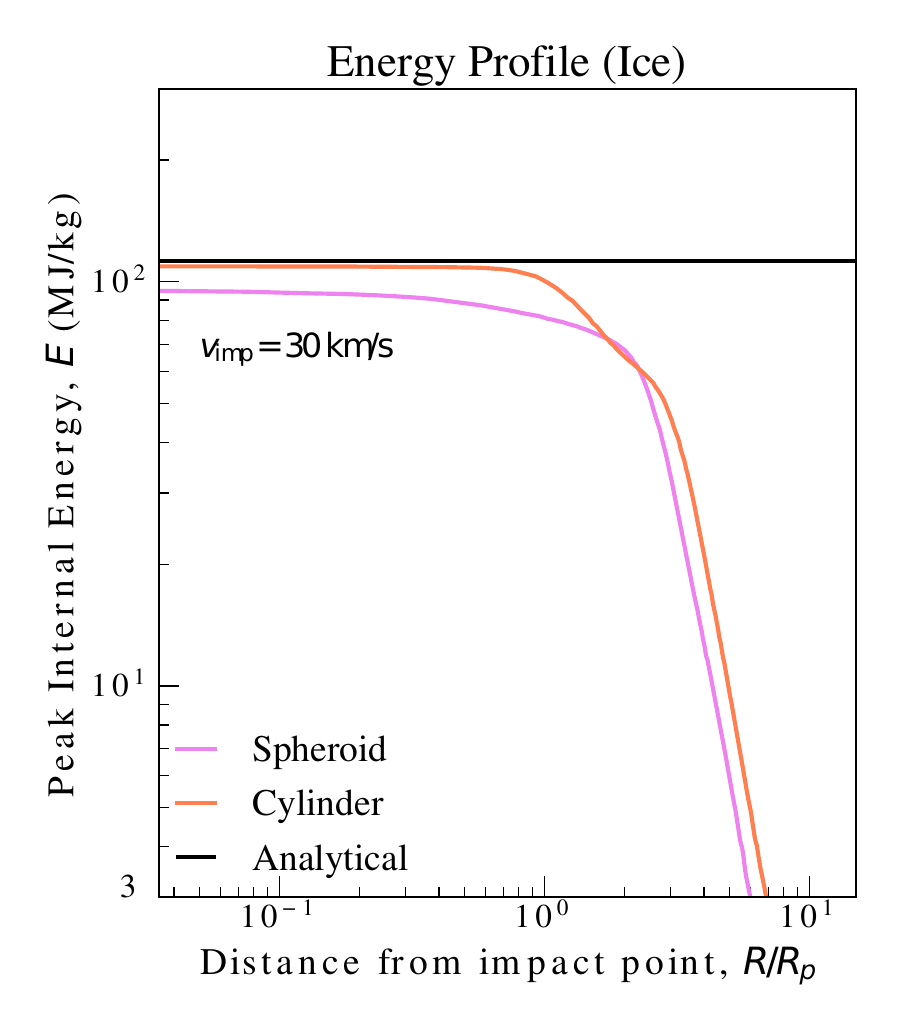}
    \caption{{Shock internal-energy field for spheroid and cylinder projectiles with $\vimp$ 30\,km/s. For the projectiles, their radii and height $R_{\rm p}$ are 500\,m. We note that the cylinder projectile with a slab target differs from the collision between cylinders in Fig.~\ref{fig:shocktube}.} The black line is given in Eq.~(\ref{eq:anal_iso_ene}) with Eq.~(\ref{eq:anal_iso_up}). }
    \label{fig:stamp}
\end{figure}

We also investigate the dependence of shock fields on the projectile shapes (Fig.~\ref{fig:stamp}). For cylinders, $E$ obtained from the simulations is in good agreement with that given by Eq.~(\ref{eq:anal_iso_ene}). For spherical bodies, $E$ obtained from the simulations is slightly smaller than Eq.~(\ref{eq:anal_iso_ene}), because the one dimensional assumption is not so valid. However, the difference is minor.

\subsubsection{Decay region}\label{subsubsec:decay}
We focus on the region distant from the impact point. The shock field for $E$ and $P$ decays as a function of distance, in a similar way to a blast wave induced by a strong explosion. For spherical symmetry, a self-similar solution determines the shock field. Therefore, the dimensionless length $\xi$ then gives the solutions \citep[e.g., ][]{Landau_fluid}, where
\begin{eqnarray}\label{eq:xi}
    \xi = \left(\frac{\rho_0}{{\cal K}_{\rm imp}t^2} \right)^{\frac{1}{5}}R_{\rm c}.
\end{eqnarray}
Here $R_{\rm c}$ is the distance from the center of the isobaric core, whereas the distance is measured from the explosion point in the blast wave problem, and ${\cal K}_{\rm imp}$ is the impact kinetic energy. Projectiles are much smaller than targets, and therefore
\begin{eqnarray}\label{eq:impact_ene}
    {\cal K}_{\rm imp}\ =\ \frac{2\pi}{3}R_{\rm p}^3\rho_0v_{\rm imp}^2.
\end{eqnarray}
where $R_{\rm p}$ is the radius of the projectile. 

The peak values for $E$ and $P$ are achieved behind the shock wave location $R_{\rm s}$. The shock wave velocity $\vs$ is given by
\begin{eqnarray}
    \vs &=& \dot{R_{\rm s}} =\frac{2\xi_{\rm s}^{5/2}{\cal K}_{\rm imp}^{1/2}}{5\rho_0^{1/2}R_{\rm s}^{3/2}} ,\label{eq:shock_velo}
\end{eqnarray}
where 
\begin{eqnarray}
    \xi_{\rm s} &=& \left(\frac{\rho_0}{{\cal K}_{\rm imp}t^2} \right)^{\frac{1}{5}}R_{\rm s}.\label{eq:xi_shock}
\end{eqnarray}
Here we assume the EoS given in Eq.~(\ref{eq:simple_eos2}). Using the Hugoniot curve given in Eqs.~(\ref{eq:anal_iso_shockvelo})–-(\ref{eq:anal_iso_ene}) with Eqs.~(\ref{eq:impact_ene}) and (\ref{eq:shock_velo}), we have the peak pressure and internal energy with $R_{\rm c}$ instead of $R_{\rm s}$:
\begin{eqnarray}
    P&=& \frac{16\pi \xi_{\rm s}^5\rho_0\vimp^2}{75\s}\left(\frac{R_{\rm c}}{R_{\rm p}} \right)^{-3},
    \label{eq:point-pressure}\\
    E &=& \frac{8\pi \xi_{\rm s}^5\vimp^2}{75\s^2}\left(\frac{R_{\rm c}}{R_{\rm p}} \right)^{-3}. \label{eq:point-intene}
\end{eqnarray}

To obtain the constant $\xi_{\rm s}$, we use the energy conservation. We consider that the initial thermal energy of colliding bodies is much smaller than the impact energy, ${\cal K}_{\rm imp}$. 
The total energy of the inner region of the target material bounded by the shock wave then corresponds to ${\cal K}_{\rm imp}$:  
\begin{eqnarray}\label{eq:ene_conserv_dimen}
    \int^{R_{\rm s}}_0\left(\frac{1}{2}\rho \up^{2} + \rho E \right)R_{\rm c}^2dR_{\rm c}  = {\cal K}_{\rm imp}. 
\end{eqnarray}
Equation~(\ref{eq:ene_conserv_dimen}) can be rewritten as 
\begin{eqnarray} 
\label{eq:ene_conserv_nondim}
    \xi_{\rm s}^5\frac{16\pi}{25}\int^1_0\left(\frac{1}{2}G V^{2} + \frac{Z}{2(\s-1)}\right)\chi^4d\chi  =1,
\end{eqnarray}
where $G, V$, and $Z$ are functions of $\chi$, as given in Appendix.~\ref{appendix}. 
Therefore, $\xi_{\rm s}$ is given by Eq.~(\ref{eq:ene_conserv_nondim}) with Eqs.~(\ref{eq:exact_nondim})-(\ref{eq:exponent_nondim}), and then $\xi_{\rm s}=1.09$ and 1.19 for ice ($\s=1.26$) and basalt ($\s=1.4$), respectively. 

We assume that the internal energy is given by the harmonic mean of the shock internal energies in the isobaric core $E_{\rm core}$ (Eq.~\ref{eq:anal_iso_ene}) and the decay region $E_{\rm decay}$ (Eq.~\ref{eq:point-intene}).
\begin{eqnarray}\label{eq:harmonic_mean}
    E(R)&=&E_{\rm core}\ \quad {\rm for}\ R\le R_0, \nonumber\\
    E(R)&=&\left(\frac{1}{E_{\rm core}} + \frac{1}{E_{\rm decay}} \right)^{-1} \quad {\rm for}\ R>R_0,\label{eq:ana_ene} 
\end{eqnarray}
where $R_0$ is set at the center of the isobaric core. 
To apply $E_{\rm decay}$ given in Eq.~(\ref{eq:point-intene}), we set $R_{\rm c} = R -R_0$ and $R_0 =1.8\,R_{\rm p}$ for ice and basalt, respectively. 
Although the solution depends on materials $E\propto \xi_{\rm s}^5\s^{-2}$ (see Eq.~\ref{eq:point-intene}), the values of $\s$ and $\xi_{\rm s}$ are similar, namely $(\s,\xi_{\rm s})$ = (1.26,\ 1.09) and (1.4,\ 1.19) for ice and basalt, respectively. Therefore the solution is almost independent of the type of material. 

\begin{figure}[htbp]
    \centering
    \includegraphics[width=7cm]{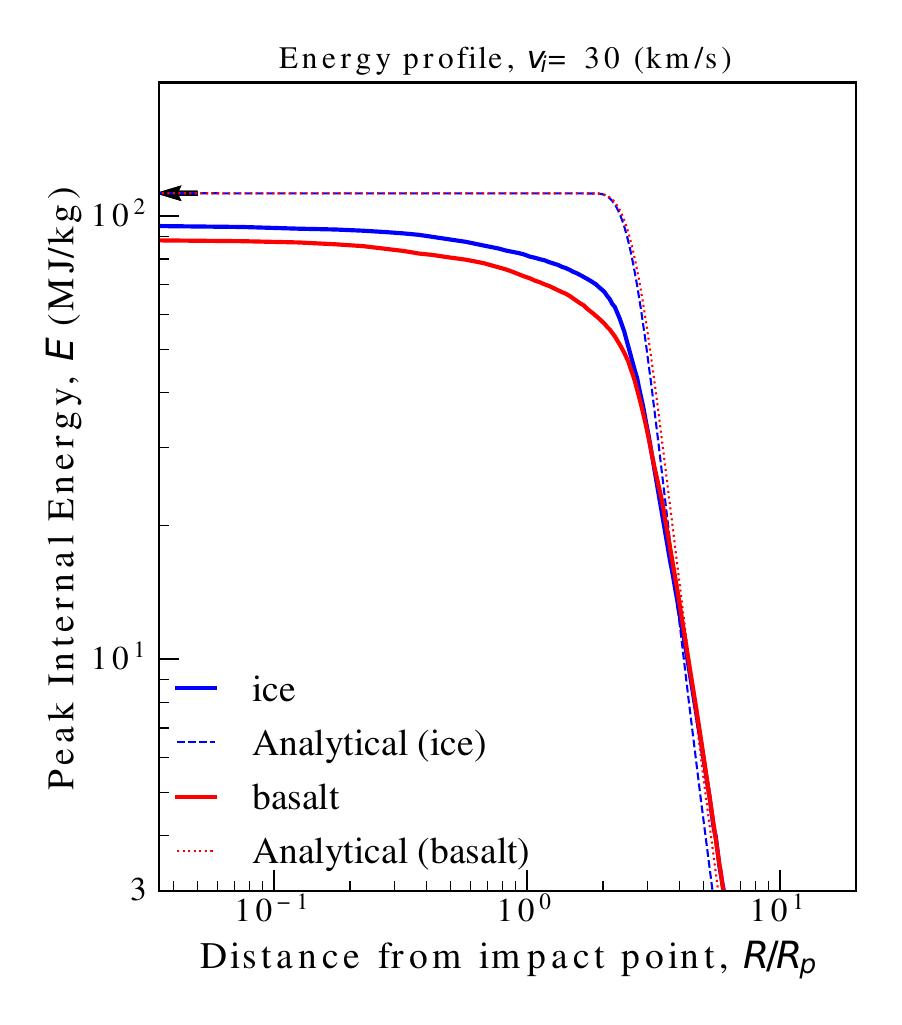}
    \caption{Shock internal-energy field for ice (blue) and basalt (red) bodies with $\vimp$=30\,km/s. Solid curves are the results of simulations. Dotted lines are the analytic solution given by Eq.~(\ref{eq:ana_ene}) with $R_0=1.8R_{\rm p}$. For reference, the black arrow is the isobaric-core solution in Eq.~(\ref{eq:anal_iso_ene}) with Eq.~(\ref{eq:anal_iso_up}), which is independent of the type of the material.
 }
    \label{fig:int_ene_anal}
\end{figure}

We show the solutions for each material as dotted lines in Fig.~\ref{fig:int_ene_anal} and compare them with simulation results. Figure~\ref{fig:int_ene_anal} corresponds to the shock internal-energy field for impacts between icy or basalt bodies with 30\,km/s. Because of the similarity between the values of $\xi_{\rm s}$ and $\s$, the solutions for the two different materials mostly reproduce the same value. Also, the solutions are mostly consistent with the simulations at $R\lesssim2R_{\rm p}$ as discussed in Sect.~\ref{subsubsec:frat}. 

\subsection{Analytical solution for shock pressure}\label{subsec:anal_pre}

\begin{figure}[htbp]
\centering
  {\includegraphics[width=8cm]{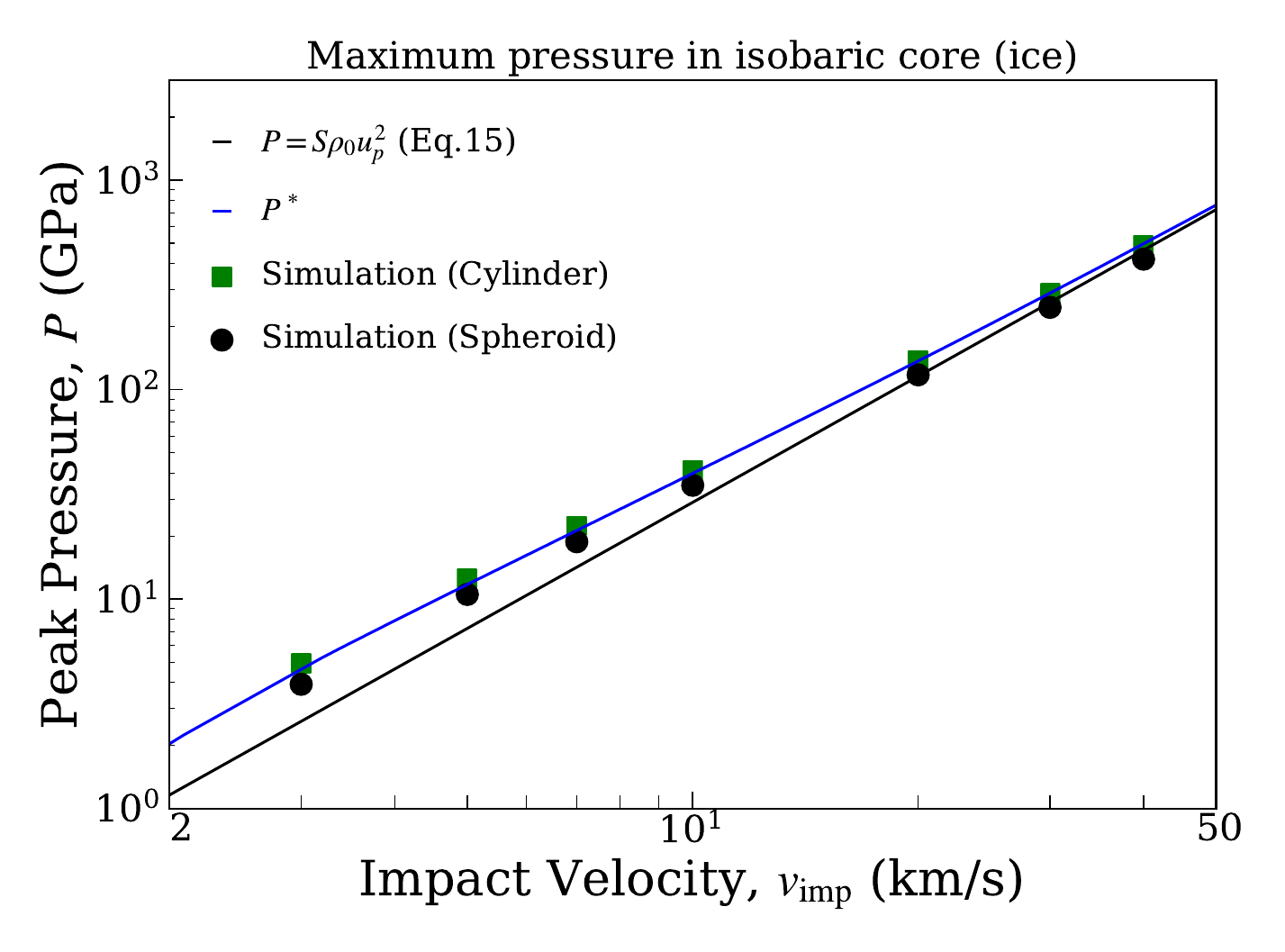}}
  {\includegraphics[width=8cm]{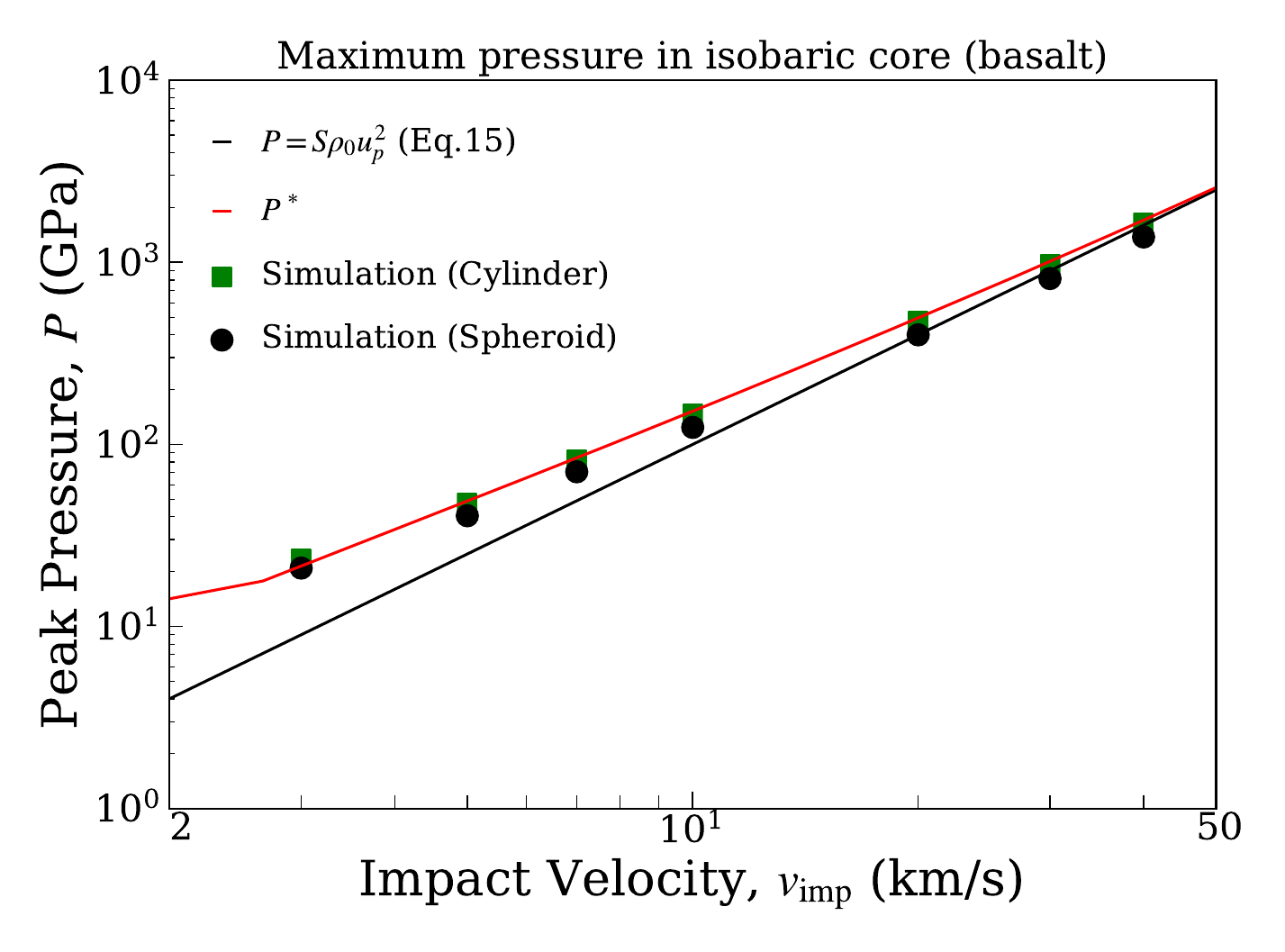}
}
\caption{Comparison between analytical solutions and simulation results. The peak pressures of these plots are recorded in the nearest tracer particle to the impact point. The shape and color of the markers in the plots correspond to projectile shape: cylinder (green square) and spheroid (black circle). {The simulation with the cylinder projectile is the same as that shown in Fig.~\ref{fig:stamp}. }
Solid lines are illustrated by Eq.~(\ref{eq:anal_iso_pre}) and $P^*$ (Sect.~\ref{subsec:anal_pre}).
\label{fig:pre_isobaric}
}
\end{figure}
\begin{figure}[htbp]
\centering
   {\includegraphics[width=7cm]{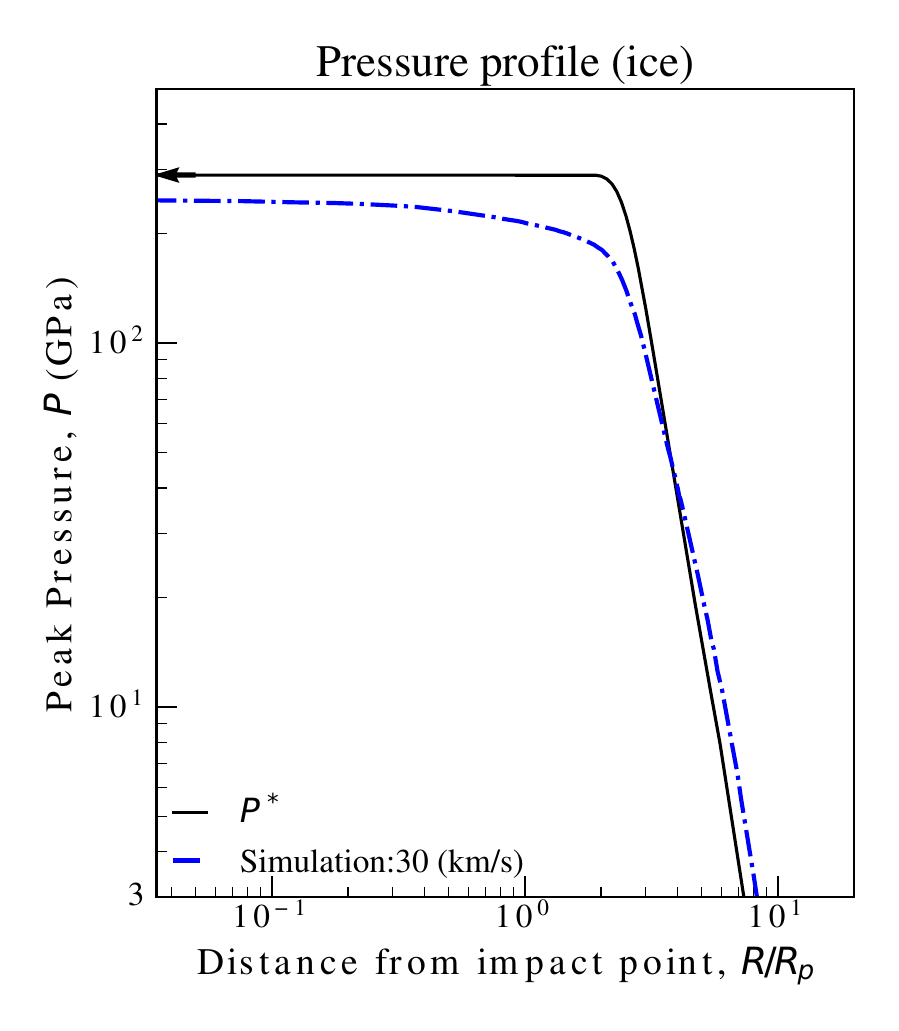}}
  {\includegraphics[width=7cm]{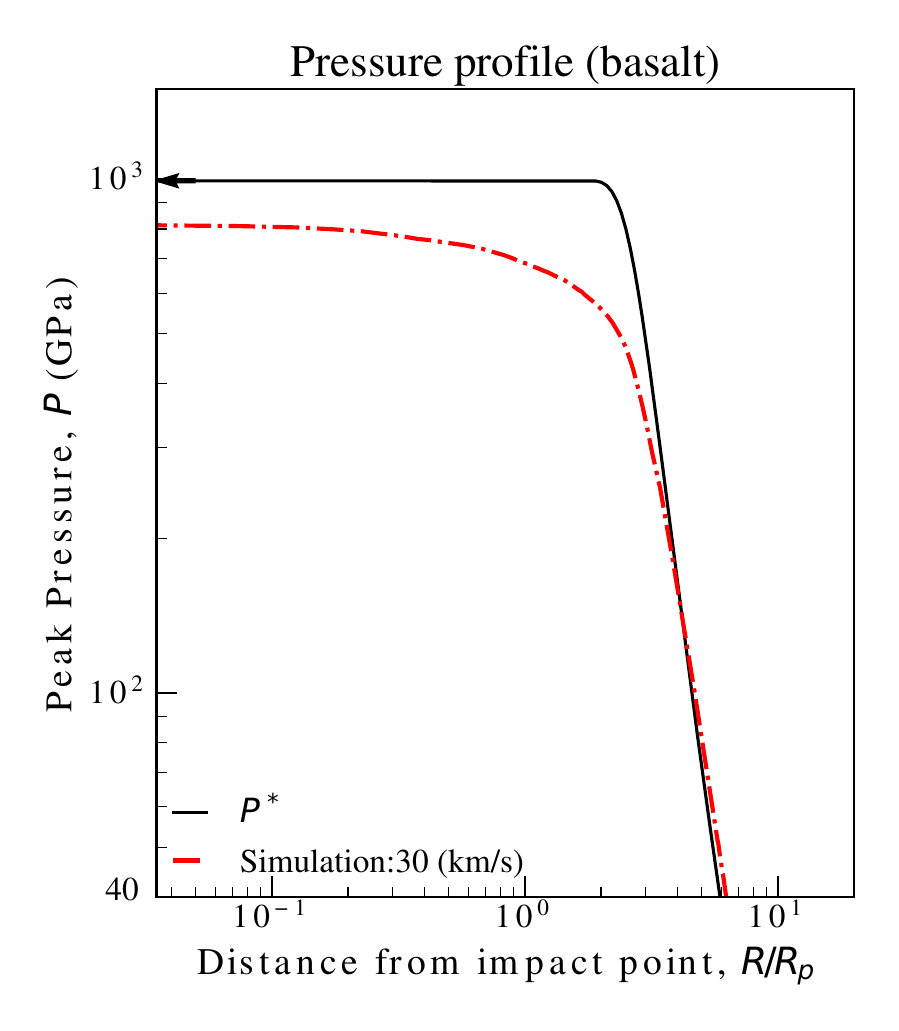}
}
\caption{Shock pressure fields in an impact event with impact velocity $\vimp=30\,$km/s for ice (a) and basalt (b) obtained from iSALE simulations {(long-dashed lines)} and from the analytical solution (solid lines). For reference, the black arrow indicates the analytic solution in the isobaric core. 
\label{fig:prepro_anal}
}
\end{figure}

As shown in Sect.~\ref{subsubsec:frat} and \ref{subsubsec:decay}, the internal energy is reproduced by the analytic solution with the EoS of Eq.~\ref{eq:simple_eos2}. However, the analytic solution underestimates the pressure for $P \la 20$\,GPa (ice) and 300\,GPa (basalt). This is caused by the importance of $P_{\rm c}$ for low pressure (see Fig.~\ref{fig:hugo_cold}). 
Therefore, the additional modification is needed to compensate for this underestimation. 

We consider $E$ --as derived in Eqs.~(\ref{eq:anal_iso_ene}) and (\ref{eq:point-intene})-- to be accurate. We then recalculate the density $\rho^*$ corresponding to $E$ from the Hugoniot curve given by ANEOS {(see lower two panels in Fig.~\ref{fig:hugo_cold}).} We obtain $P^*$ corresponding to $\rho^*$ from the Hugoniot curve {(upper two panels in Fig.~\ref{fig:hugo_cold}).} The shock pressure is thus given by $P^*$. 

Figure~\ref{fig:pre_isobaric} shows the pressures at isobaric cores as a function of impact velocity. Equation~(\ref{eq:anal_iso_pre}) is in good agreement with the simulation results for the high velocity, while Eq.~(\ref{eq:anal_iso_pre}) underestimates for low-velocity region. The modified pressure $P^*$ reproduces the simulation results even for low velocities. 

The agreement between the analytical solution and simulation result for $E$ and $P^*$ indicates the validity of the assumption. The shock wave is mainly driven by the thermal terms. The internal energy is accurately estimated from the simple EoS of Eq.~(\ref{eq:simple_eos2}). The pressure and density are determined by the internal energy induced by a shock wave, and are obtained from the values corresponding to the internal energy according to the Hugoniot curve. 

According to the analytic solution for the internal energy, we assume that the pressure is given by the harmonic mean of $P^*_{\rm core}$ and $P^*_{\rm decay}$. 
\begin{eqnarray}\label{eq:harmonic_mean_pre}
    P^*(R)&=&P^*_{\rm core}\ \quad {\rm for}\ R\le R_0, \nonumber\\
    P^*(R)&=&\left(\frac{1}{P^*_{\rm core}} + \frac{1}{P^*_{\rm decay}} \right)^{-1} \quad {\rm for}\ R>R_0,\label{eq:ana_pre} 
\end{eqnarray}
where $P^*_{\rm core}$ and $P^*_{\rm decay}$ are the analytic solution for the isobaric core and the decay region, respectively, and 
the setting of $R_0$ is the same as the internal energy. 
Figure~\ref{fig:prepro_anal} shows the shock pressure fields obtained from simulations and the analytical solution for icy and basaltic bodies with $\vimp=30$\,km/s. The analytical solution $P^*$ reproduces the shock pressure field from the isobaric core to the decay region.

\section{Angle dependence}\label{sec:discussion}
The previous section focus on the vertical shock fields in previous sections. 
In this section, we discuss the shock fields along other directions and investigate the dependence of shock pressure fields on the angles from the impact point.

\begin{figure}[htbp]
\centering
    \includegraphics[width=7cm]{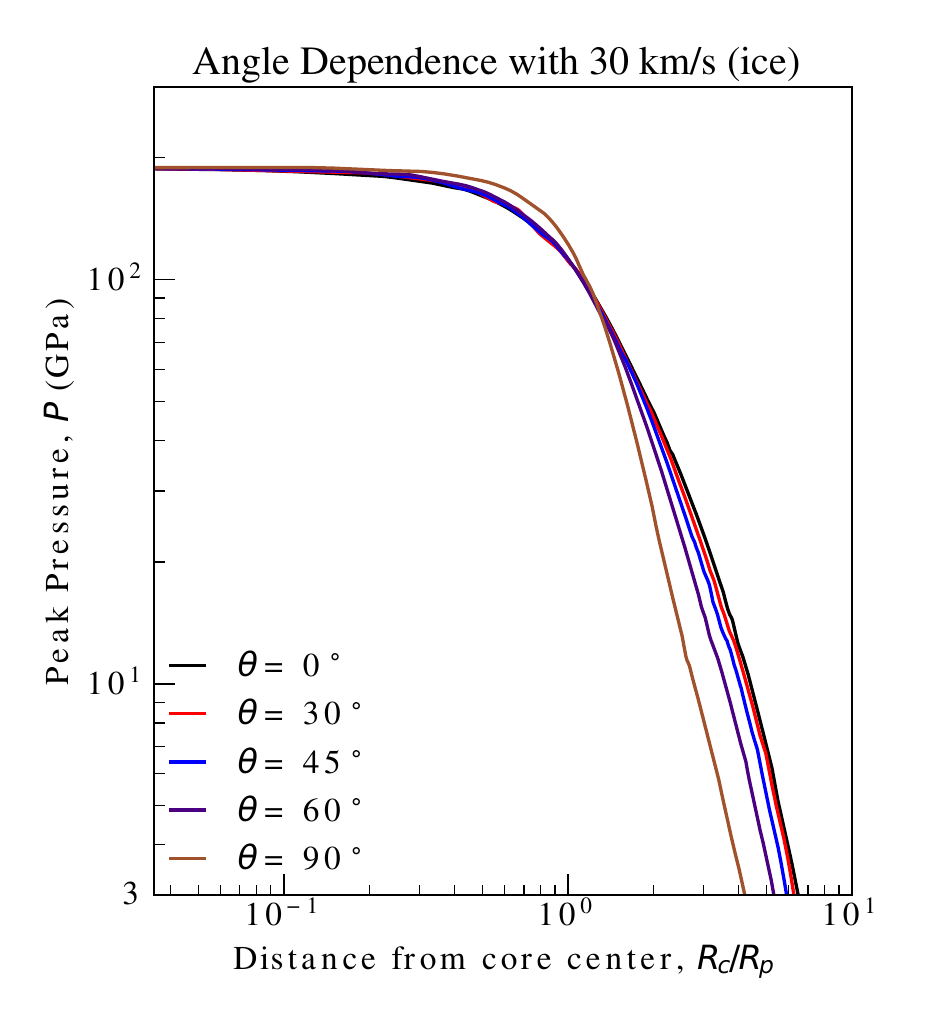}
    \includegraphics[width=7cm]{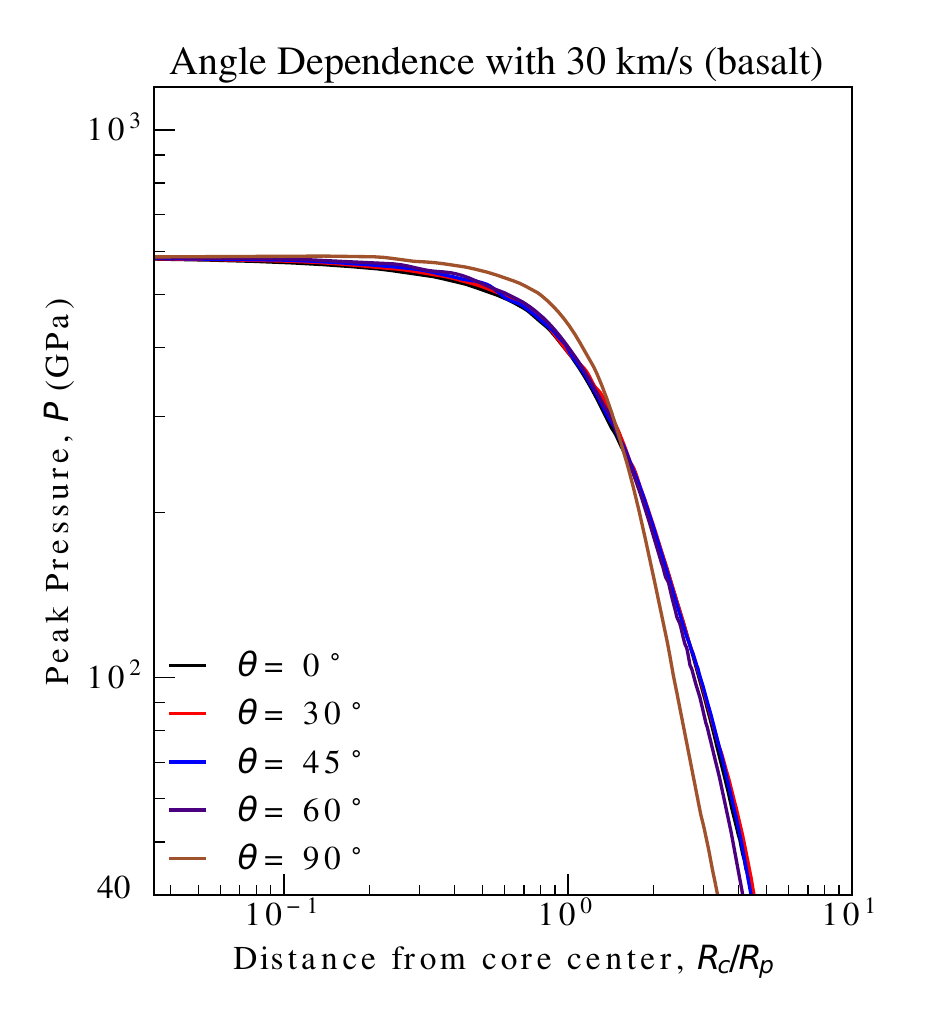}
\caption{{Dependence of shock pressure on the angle of the shock field to the vertical direction.} The figure shows shock pressure fields for ice and basalt bodies as a function of the distance from the center of isobaric core $R_{\rm c}$ (Sect.~\ref{subsubsec:decay}). Each color corresponds to {a different angle} measured from the core center and $\theta=0^\circ$ is defined as the vertical direction. 
\label{fig:angle_depend}
}
\end{figure}

Figure~\ref{fig:angle_depend} shows the shock pressure fields with $\vimp=30$\,km/s. We note that the horizontal value $R_{\rm c}$ is not the distance from the impact point $R$, but the distance from the center of the core. We set the center of core at $z = -1.8R_{\rm p}$, which is same as $R_0$ (see Sect.~\ref{subsubsec:decay}). It should be noted that the angle $\theta$ is measured from the vertical direction; that is $\theta = 0$ is for the vertical shock fields. 
As shown in Fig.~\ref{fig:angle_depend}, the pressure fields are almost independent of $\theta\la 60^\circ$, while the angle dependence is strong for $\theta\gtrsim60^\circ$. 
The interference between rarefaction and shock waves may modify the pressure field especially for $\theta \ga 90^\circ$. 
In addition, for large angles $\theta\gtrsim60^\circ$, the shock wave propagation deviates from the isotropic expansion from the core assumed for our analysis. However, the relative difference between pressures from the analytical solution and the numerical result at $R_{\rm c}=3R_{\rm p}$ is within 30\% if $\theta \le 60^\circ$ for both materials, but is within 10\% for basalt (Fig.~\ref{fig:angle_depend}).

In Sect.~\ref{subsubsec:decay}, our solutions based on the assumption of spherical symmetry are valid for the shock fields orientated in the vertical direction $\theta=0^\circ$. 
Figure~\ref{fig:angle_depend} shows that our assumption of the spherical symmetry is valid for the shock fields with small angles, such as $0\le\theta\le60^\circ$ for ice and basalt. 

\section{Atmosphere formation: Impact vaporization}\label{sec:atmos}
\subsection{The amount of vapor due to an impact}\label{subsec:vaporamount}

\begin{figure}[htbp]
\centering
    \includegraphics[width=8cm]{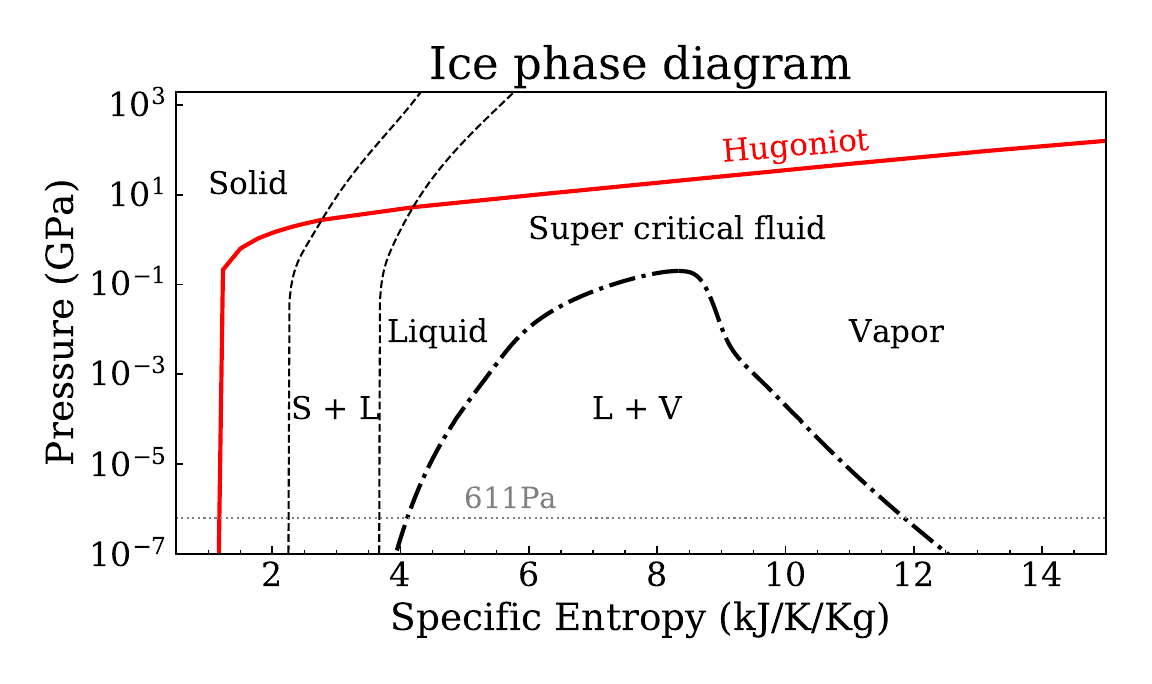}
    \includegraphics[width=8cm]{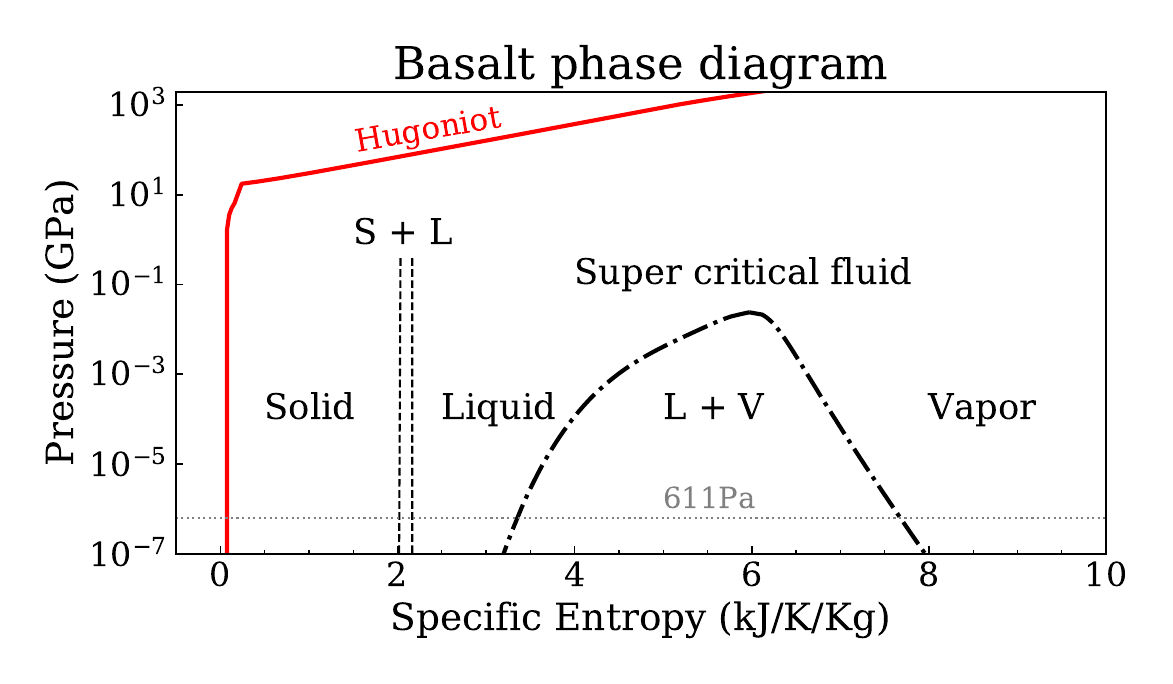}
\caption{Phase diagram for water ice and basalt from the ANEOS \citep{2001Sci...294.1326T,2005..basalt..pierazzo} with the Hugoniot curves (red line). {The dotted line and dash-dotted lines correspond to the phase boundary for solid--liquid and liquid--vapor, respectively. }
\label{fig:phasediagram}
}
\end{figure}

\begin{figure}[htbp]
\centering
 \includegraphics[width=8cm]{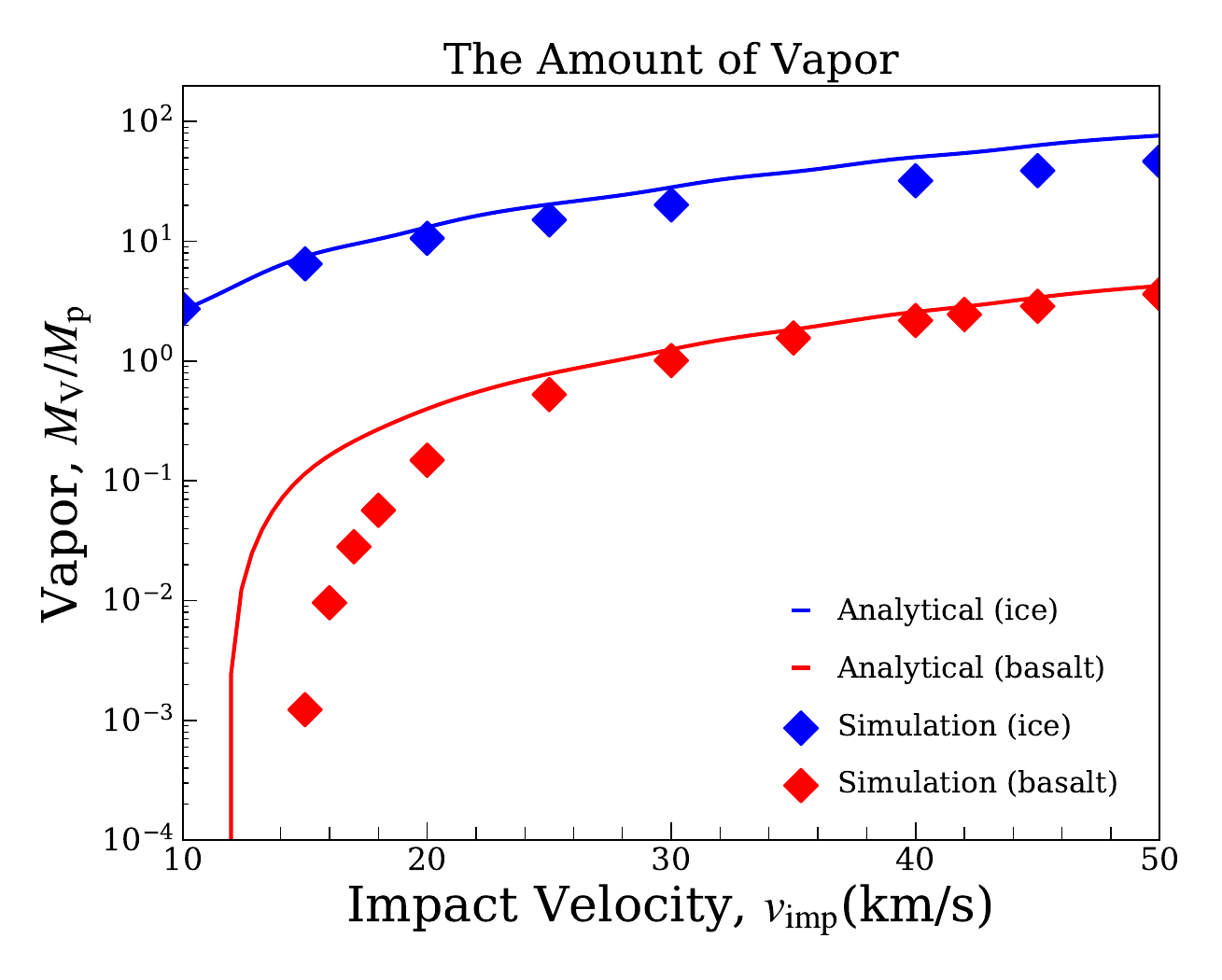}
\caption{Scaled mass of water ice and basalt that have been vaporized. These plots represent simulation results and lines are given by our solutions. The mass of vapor $M_{\rm V}$ includes both the mass of the target and the projectile. 
\label{fig:vapor_sim}
}
\end{figure}

The shock wave in impact events can generate an extremely high temperature
field, and so this event may shape planetary atmospheres. The accretion
process contributes to the growth of terrestrial atmospheres
\citep[e.g.,][]{Benlow_Meadows1977,Lange_Ahrens_1982Icar}. The amount of
vapor production is derived from the shocked state because the
subsequent process leading to shock compression can be assumed as the
isentropic expansion \citep{sugita+2002}; therefore, the shock entropy
field determines the amount of vapor \citep{1977iecp.symp..639A}. For
that reason, the isentropic expansion after the shock compression is
illustrated by the vertical line from a point on the Hugoniot curve to
the horizontal direction in Fig.~\ref{fig:phasediagram}. The final
pressure during isentropic expansion depends on the ambient
pressure. In Fig.~\ref{fig:vapor_sim}, we set the final pressure
$P_{\rm F}$ at the triple point pressure of $\rm H_2O$, $P_{\rm
F}=611$\,Pa, which is the same as in \citet{2011Icar..214..724K}. For material
that reaches the intermediate regions between liquid and vapor, the mass
of the partial vapor is given by the lever arm rule \citep[e.g.,
][]{2011Icar..214..724K}.

Our analytic solutions allow us to obtain the entropy field using the ANEOS Hugoniot relation, $S=S(E)$. From this, we can derive the amount of vapor as a function of impact velocity (Fig.~\ref{fig:vapor_sim}). 

\begin{figure}[htbp]
\centering
    \includegraphics[width=8cm]{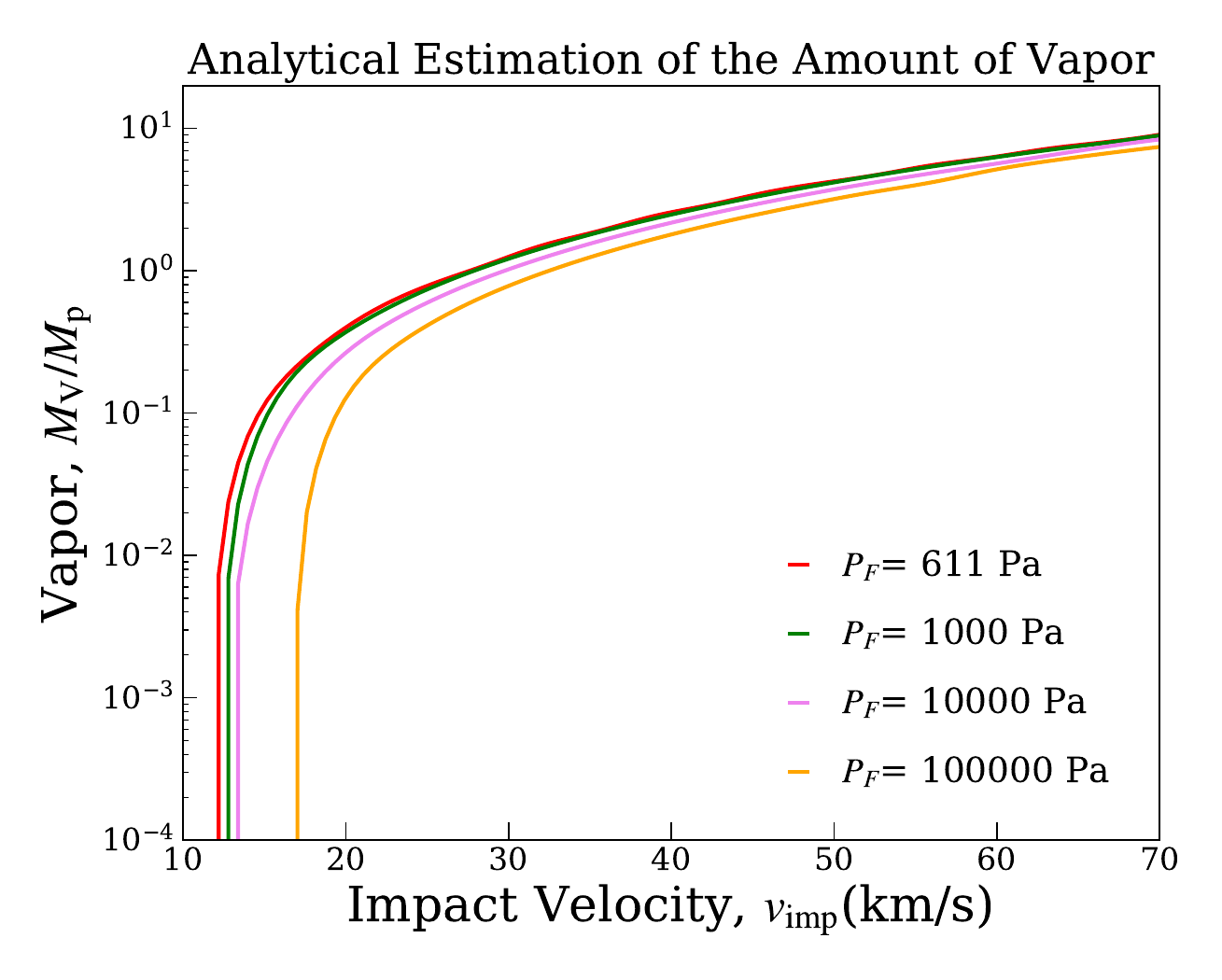}
\caption{Analytical estimation of the mass of the vapor production of basaltic bodies for target and projectile materials, scaled by the projectile mass. Each color corresponds to {different final pressure $P_{\rm F}$.} 
\label{fig:vapor_analytical}
}
\end{figure}

The analytic solutions reproduce the consistent value with simulation results (Fig.~\ref{fig:vapor_sim}). 
The analytical solution overestimates the amount of vapor compared to simulation results for basaltic bodies with low impact velocities of $\vimp\la20\,$km/s, for which vaporization mainly occurs in isobaric cores. Although our analytic solution accurately represents the peak pressure in the isobaric core (Fig.~\ref{fig:pre_isobaric}), the shock pressure in the core slightly decreases with distance (Fig.~\ref{fig:prepro_anal}). Our analytic solution ignores this slight decline in the isobaric core, and gives the amount of vapor according to the isobaric-core solution. The slight overestimation around the edges of isobaric cores leads to the overestimation in the amounts of vapors for basalt bodies with low velocities $\vimp\la20\,$km/s (Fig.~\ref{fig:vapor_sim}). However, our analytic solution is useful for estimating the amount of impact vapor. Figure~\ref{fig:vapor_analytical} illustrates our analytical estimation of the amount of vapor for an impact on a basaltic planet with an ambient pressure of $P_{\rm F}$. The amount of vapor is the increase function of impact velocity. If impact velocities are higher than $20 \,$km/s, the dependence on $P_{\rm F}$ is small.

It should be noted that we obtain the amount of vapor under the assumption of fluid dynamics. Other heating effects, such as shear heating, may influence the shocked state especially below $\vimp \la$\,12\,km/s \citep{Manske+2022}. {However, our analytical solution is in good agreement with the numerical results for the high-speed impact region at $\vimp\ga25$\,km/s, with the relative error between the two results being less than 60$\%$.}

\subsection{Reformation of planetary atmospheres during accretion process in a giant impact}\label{subsec:formation_atm}

In the final formation stage of the terrestrial planets of the Solar System, mars-sized protoplanets formed and terrestrial planets then formed via collisions between these protoplanets, called giant impacts, which were induced by the long-term chaotic orbital instability \citep[e.g.,][]{Chambers+98}. 

In a giant-impact event, the planetary bodies receive violent impact energies, and then eject some of their material \citep[e.g., ][]{Shuvalov+2014,Stewart+2014,Kurosaki+2023}. The planetary atmospheres are easily removed by the impact event \citep{Kurosaki+2023}. On the other hand, the secondary accretion of solid giant-impact ejecta may cause reformation of the atmospheres. In this section, we estimate the formation of the atmospheres through the accretion process. 

A giant impact ejects a certain amount of small bodies, which amount to approximately 10 \% of the mass of the colliding protoplanets \citep{genda15}. 
The ejecta from a giant impact finally distribute as a disk around the host star. The total mass of the giant-impact ejecta mainly decreases due to the collisional cascade among the giant-impact ejecta with the blow-out of micron-sized or smaller bodies \citep{Kobayashi+2019}. However, some of the giant-impact ejecta are accreted onto the parent planet. 

Considering a planet with mass {$M_{\rm planet}$} around a host star of solar mass, the relative velocity $v_{\rm rel}$ between the planet and the ejecta is roughly given by 1--$3 v_{\rm esc}$, where $v_{\rm esc}$ is the escape velocity of the planet \citep{genda15}. Here we set to $v_{\rm rel} = \sqrt{3}v_{\rm esc}$. 
The accretion timescale for the reduction of the giant-impact ejecta is estimated as 
\begin{eqnarray}\label{eq:acc_time}
 \tau_{\rm acc}\approx 3\times10^8\left(\frac{M_{\rm planet}}{M_{\oplus}}\right)^{\frac{1}{3}}\left(\frac{a}{1\rm AU}\right)^4\left(\frac{\rho}{{3\,\rm g/cm}^{3}}\right)^{-\frac{5}{6}}\ \rm yr,     
\end{eqnarray}
where $a$ is the semimajor axis of the planet, $\rho$ is the solid density of the planet and the ejecta, and $M_{\oplus}$ is the Earth mass. 

On the other hand, the timescale of the collisional cascade is 
characterized by the mass of largest ejecta $m_{\rm c}$, which is 
estimated to \citep{kobayashi_tanaka10,Kobayashi+2019}, 
\begin{eqnarray}
 \tau_{\rm cc} &\approx& 1\times10^3\left(\frac{M_{\rm ej}}{M_{\oplus}}\right)^{-1}\left(\frac{a}{1\rm AU}\right)^4
\nonumber
\\
&& \times
\left(\frac{m_{\rm c}}{10^{21}\rm g}\right)^{0.64}\left(\frac{M_{\rm planet}}{M_{\oplus}}\right)^{0.12}\left(\frac{\rho}{3\,{\rm g/cm}^{3}}\right)^{0.61}\ \rm yr,
\label{eq:cas_time}
\end{eqnarray}
where $M_{\rm ej}$ is the total mass of the giant-impact ejecta.

For a giant impact of protoplanets, $M_{\rm ej}$ is typically given by $\sim 0.1 M_{\rm planet}$. For $M_{\rm planet} = M_\oplus$, a giant impact gives $M_{\rm ej} \approx 0.1 M_{\oplus}$, resulting in $\tau_{\rm cc} \ll \tau_{\rm acc}$ (compare between Eqs. \ref{eq:acc_time} and \ref{eq:cas_time}). The giant-impact ejecta mass $M_{\rm ej}$ thus decreases via the collisional cascade. Once $M_{\rm ej} \lesssim 10^{-5} M_\oplus$, $\tau_{\rm cc}$ is longer than $\tau_{\rm acc}$. The giant-impact ejecta are then mainly accreted onto the planet. Therefore, the accretion mass of giant-impact ejecta $M_{\rm ej,ac}$ is estimated from $\tau_{\rm acc} = \tau_{\rm cc}$ as, 
\begin{equation}\label{eq:mass_eject}
  M_{\rm ej,ac} \approx 3 \times 10^{-6} 
   \left(\frac{m_{\rm c}}{10^{21}\rm g}\right)^{0.64}
   \left(\frac{M_{\rm planet}}{M_{\oplus}}\right)^{-1.21}
   \left(\frac{\rho}{3\,{\rm g/cm}^{3}}\right)^{1.44}\ M_{\rm p}. 
\end{equation}
Interestingly $M_{\rm ej,ac}$ is independent of $a$, because the dependence of $\tau_{\rm acc}$ and $\tau_{\rm cc}$ on $a$ is the same. $M_{\rm ej,ac}$ is larger for larger $m_{\rm c}$. {The accretion mass $M_{\rm ej,ac}$ decreases with $M_{\rm planet}$. } 

\begin{figure}[htbp]
\centering
    \includegraphics[width=8.5cm]{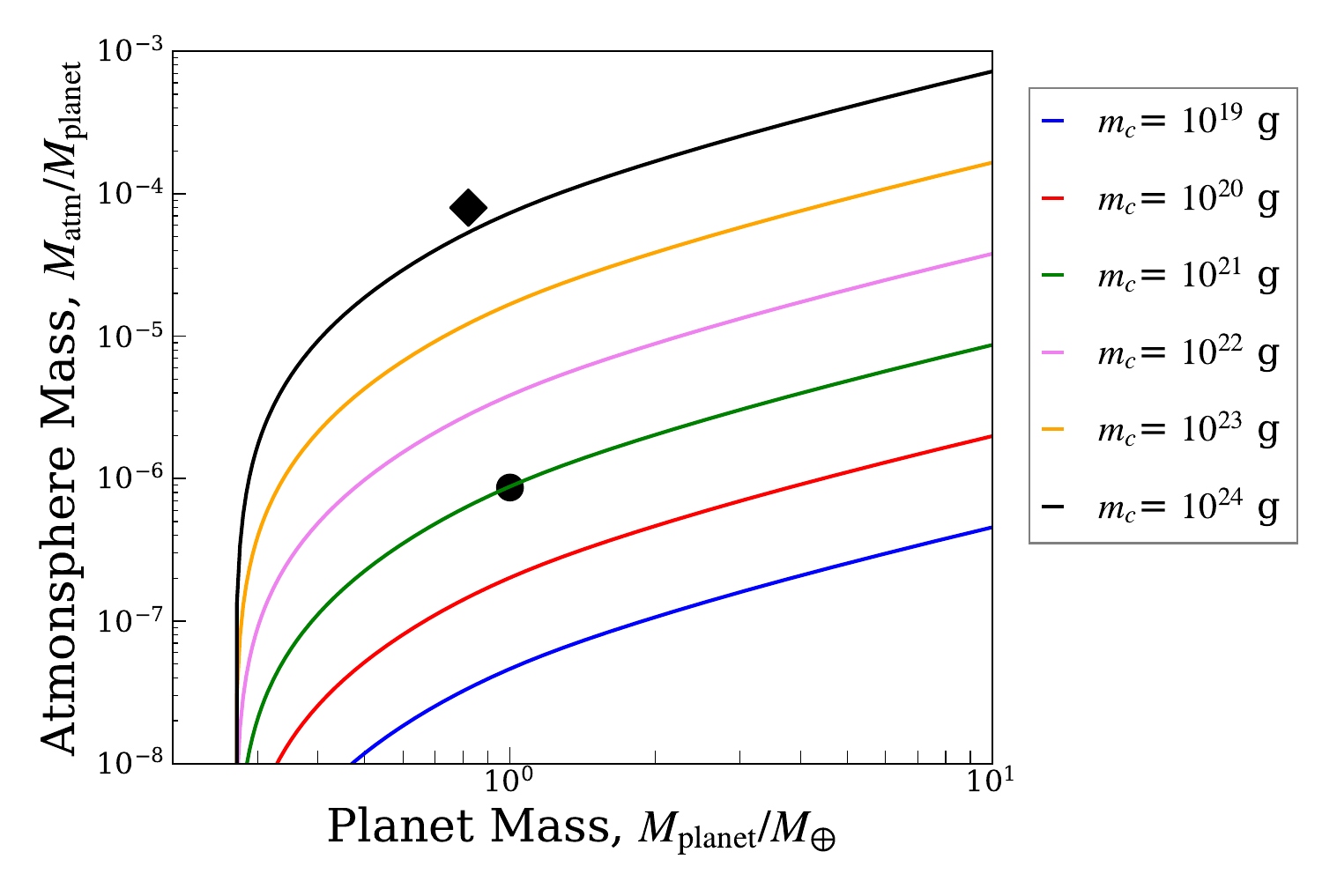}
\caption{Atmosphere mass $M_{\rm atm}$ produced by the accretion process following a giant impact, as a function of the planetary mass $M_{\rm planet}$. {Each color corresponds to the mass of the largest ejecta $m_{\rm c}$}. The circle and diamond represent the atmospheric masses of Earth and Venus, respectively. 
\label{fig:atmosphere}
}
\label{fig:atm}
\end{figure}

Using $M_{\rm ej,ac}$ with the impact velocity given by {$\vimp\simeq \sqrt{v^2_{\rm rel}+v^2_{\rm esc}}=2v_{\rm esc}$}, we calculate the atmosphere mass $M_{\rm atm}$ composed of vapor caused by the impacts with the giant-impact ejecta for $P_{\rm F}=10^5$\,Pa (see Fig.~\ref{fig:atm}). Again, we note that the choice of $P_{\rm F}$ is unimportant here (see Figure~\ref{fig:vapor_analytical}). For $M_{\rm planet} = M_\oplus$, $v_{\rm imp} \approx 22 \,$km/s, resulting in $M_{\rm atm} \approx 0.3 M_{\rm ej,ac}$. 
As shown in Fig.~\ref{fig:atm}, the atmospheric mass of Earth is reproduced by impact vaporization with $m_{\rm c} \ga 10^{21}$\,g, while Venus's atmosphere needs $m_{\rm c} \ga 10^{24}$\,g. 
{While the accretion mass $M_{\rm ej,ac}$ decreases with $M_{\rm planet}$ (Eq.~\ref{eq:mass_eject}), the vapor mass significantly depends on $v_{\rm imp}$ (Fig.~\ref{fig:vapor_analytical}) and thus $M_{\rm atm}$ increases with $M_{\rm planet}$ because of the dependence of $v_{\rm imp}$ on $M_{\rm planet}$. }
Post-impact atmospheres become composed of $\rm H_2O$ and $\rm CO_2$ via chemical evolution on a timescale of $10^6$ years \citep{Lupu+14}.
The substantial evolution mainly reduces the atmosphere mass. The giant-impact ejecta with $m_{\rm c} \ga 10^{24}$\,g are likely to reproduce planetary atmospheres in the Solar System.


We note that $m_{\rm c}$ depends on the collisional parameters of
giant impacts. According to collisional simulations with collisional parameters given by $N$-body simulations for giant-impact stages, typical giant impacts result in $m_{\rm c} \gtrsim 10^{24}$\,g \citep{genda15}. 
In addition, planets resulting from $N$-body simulations have highly eccentric orbits, which are inconsistent with the present orbits of Earth and Venus. 
The dynamical friction with giant-impact ejecta is possible to decay the eccentricities. This decay requires $\tau_{\rm cc}$, which is longer than the timescale of the dynamical friction, which means $m_{\rm c} \ga 10^{24}$\,g \citep{Kobayashi+2019}. The terrestrial planets in the Solar System may be formed from giant impacts with $m_{\rm c} \ga 10^{24}$\,g. On the other hand, if giant impacts result in smaller $m_{\rm c}$, the resultant planets may have high eccentricities and small $M_{\rm atm}$. Therefore, our scenario implies that $M_{\rm atm}$ has the correlation with orbital eccentricities. 

{
Above, we discuss the formation of secondary atmospheres subsequent to a giant impact. A primary atmosphere is obtained from a protoplanetary disk and is mainly composed of $\rm H_2$ and $\rm He$. On the other hand, the composition of the secondary atmosphere due to impacts is similar to that of atmospheres formed by reactions between a primary atmosphere and a magma ocean \citep[e.g.,][]{itcovitz+2022,kurosaki+2023b}. The primary atmosphere evolution may produce $\rm H_2O$ without experiencing a giant impact \citep[e.g., ][]{ikoma+2006,kimura+2022}. 
If a planet has an ocean, the surface ocean may influence the heating process and atmosphere stripping \citep[e.g., ][]{genda+2005,schlichting+2015,lock+2024}. Even in such complicated situations, our analytical formula may be applied to calculate the amount of vapor using the $\s$ value for oceans. However, the chemical reactions between vapors from rocky body and ocean should be considered in the subsequent evolution, and these will be addressed in a future study. 
}

\section{Summary}\label{sec:summary}
A high-speed planetary impact causes a shock wave, which is expected to occur in the late stages of planet formation. Then, the shock wave can form an extremely high-pressure field called the shock pressure field. We have investigated the structure of the shock pressure field in impact simulations with iSALE-2D using ANEOS, which is one of the most sophisticated EoS models. We have obtained two specific features of the shock field as a function of distance, the isobaric core around the impact point and the decay region distant from the impact point (Sect.~\ref{subsec:result_shockpressure}). To understand the shock fields, we have investigated the contribution of the thermal and cold terms in the solid EoS model on the Hugoniot curves. We find that the shock pressure field is determined by both the thermal and cold terms, while the cold term makes a minor contribution to the shock internal-energy field. 

We have thus used a simple EoS based only on the thermal term. To obtain analytical solutions for the shock internal-energy field, we assume the shock wave as a strong explosion with spherical symmetry. Under this assumption, we have derived analytical solutions for the shock internal-energy field (Sect.~\ref{subsec:analytical_ene}). Furthermore, we have additionally given analytical solutions for the shock pressure using the Hugoniot curve (Sect.~\ref{subsec:anal_pre}). These solutions based on the spherical symmetry are valid to vertical direction along the impact direction. 

To investigate the applicability of our solutions to other directions, we have analyzed the shock pressure field in various directions. We have found that as the angle from the vertical axis increases, the errors on the shock pressure from the vertical direction became large. {However, as discussed in Sect.~\ref{sec:discussion},} our assumption of spherical symmetry is valid for $\theta\le60^\circ$. The angle should be measured from the center of the core set properly. 

Using our solutions for the shock field, we show the amount of vapor production due to an impact event (Sect.~\ref{subsec:vaporamount}). We also estimate the planetary atmosphere mass formed by the accretion of ejecta from a giant impact (Sect.~\ref{subsec:formation_atm}). Our findings can explain the amounts of planetary atmosphere found on the terrestrial planets of the Solar System. 

\begin{acknowledgements}
{We are grateful to the anonymous referee for beneficial comments.} We thank iSALE developers, including G. Collins, K. Wünnemann, B. Ivanov, J. Melosh, and D. Elbeshausen. We used pySALEplot to analyze the result of iSALE and to make figures. We thank Tom Davison for the development of pySALEplot. Our numerical simulations were carried out on the general-purpose PC cluster at Center for Computational Astrophysics. I would like to take this opportunity to thank the "Interdisciplinary Frontier Next-Generation Researcher Program of the Tokai Higher Education and Research System". The work is supported by Grants-in-Aid for Scientific Research (21K03642, 22H01278, 22H00179) from MEXT of Japan. 
\end{acknowledgements}
\bibpunct{(}{)}{;}{a}{}{,} 
\bibliographystyle{aa} 

\begin{thebibliography}{48}
\expandafter\ifx\csname natexlab\endcsname\relax\def\natexlab#1{#1}\fi

\bibitem[{{Ahrens} \& {O'Keefe}(1972)}]{1972Moon....4..214A}
{Ahrens}, T.~J. \& {O'Keefe}, J.~D. 1972, Moon, 4, 214

\bibitem[{{Ahrens} \& {O'keefe}(1977)}]{1977iecp.symp..639A}
{Ahrens}, T.~J. \& {O'keefe}, J.~D. 1977, in Impact and Explosion Cratering: Planetary and Terrestrial Implications, ed. D.~J. {Roddy}, R.~O. {Pepin}, \& R.~B. {Merrill}, 639--656

\bibitem[{Amsden {et~al.}(1980)Amsden, Ruppel, \& Hirt}]{Amsden_etal_1980}
Amsden, A.~A., Ruppel, H.~M., \& Hirt, C.~W. 1980, SALE: a simplified ALE computer program for fluid flow at all speeds, los Alamos National Laboratory Report LA-8095

\bibitem[{{Benlow} \& {Meadows}(1977)}]{Benlow_Meadows1977}
{Benlow}, A. \& {Meadows}, A.~J. 1977, \apss, 46, 293

\bibitem[{{Bowling} {et~al.}(2020){Bowling}, {Johnson}, {Wiggins}, {Walton}, {Melosh}, \& {Sharp}}]{Bowling_Johnson_2020}
{Bowling}, T.~J., {Johnson}, B.~C., {Wiggins}, S.~E., {et~al.} 2020, \icarus, 343, 113689

\bibitem[{{Chambers} \& {Wetherill}(1998)}]{Chambers+98}
{Chambers}, J.~E. \& {Wetherill}, G.~W. 1998, \icarus, 136, 304

\bibitem[{{Croft}(1982)}]{1982GS...croft}
{Croft}, S.~K. 1982, Geological Society of America Special Paper, 190, 143

\bibitem[{{Genda} \& {Abe}(2005)}]{genda+2005}
{Genda}, H. \& {Abe}, Y. 2005, \nat, 433, 842

\bibitem[{{Genda} {et~al.}(2015){Genda}, {Fujita}, {Kobayashi}, {Tanaka}, \& {Abe}}]{genda15}
{Genda}, H., {Fujita}, T., {Kobayashi}, H., {Tanaka}, H., \& {Abe}, Y. 2015, \icarus, 262, 58

\bibitem[{{Ikoma} \& {Genda}(2006)}]{ikoma+2006}
{Ikoma}, M. \& {Genda}, H. 2006, \apj, 648, 696

\bibitem[{{Itcovitz} {et~al.}(2022){Itcovitz}, {Rae}, {Citron}, {Stewart}, {Sinclair}, {Rimmer}, \& {Shorttle}}]{itcovitz+2022}
{Itcovitz}, J.~P., {Rae}, A. S.~P., {Citron}, R.~I., {et~al.} 2022, PSJ, 3, 115

\bibitem[{Ivanov {et~al.}(1997)Ivanov, Deniem, \& Neukum}]{Ivanov_etal_1997}
Ivanov, B., Deniem, D., \& Neukum, G. 1997, International Journal of Impact Engineering, 20, 411, hypervelocity Impact Proceedings of the 1996 Symposium

\bibitem[{{Kadono} {et~al.}(2002){Kadono}, {Sugita}, {Mitani}, {Fuyuki}, {Ohno}, {Sekine}, \& {Matsui}}]{Kadono_sugita_2002GeoRL}
{Kadono}, T., {Sugita}, S., {Mitani}, N.~K., {et~al.} 2002, \grl, 29, 1979

\bibitem[{{Kimura} \& {Ikoma}(2022)}]{kimura+2022}
{Kimura}, T. \& {Ikoma}, M. 2022, Nature Astronomy, 6, 1296

\bibitem[{{Kobayashi} {et~al.}(2019){Kobayashi}, {Isoya}, \& {Sato}}]{Kobayashi+2019}
{Kobayashi}, H., {Isoya}, K., \& {Sato}, Y. 2019, \apj, 887, 226

\bibitem[{{Kobayashi} \& {Tanaka}(2010)}]{kobayashi_tanaka10}
{Kobayashi}, H. \& {Tanaka}, H. 2010, \icarus, 206, 735

\bibitem[{{Kobayashi} {et~al.}(2011){Kobayashi}, {Tanaka}, \& {Krivov}}]{kobayashi11}
{Kobayashi}, H., {Tanaka}, H., \& {Krivov}, A.~V. 2011, \apj, 738, 35

\bibitem[{{Kobayashi} {et~al.}(2010){Kobayashi}, {Tanaka}, {Krivov}, \& {Inaba}}]{kobayashi+10}
{Kobayashi}, H., {Tanaka}, H., {Krivov}, A.~V., \& {Inaba}, S. 2010, \icarus, 209, 836

\bibitem[{{Kraus} {et~al.}(2015){Kraus}, {Root}, {Lemke}, {Stewart}, {Jacobsen}, \& {Mattsson}}]{Kraus_Root_2015NatGe}
{Kraus}, R.~G., {Root}, S., {Lemke}, R.~W., {et~al.} 2015, Nature Geoscience, 8, 269

\bibitem[{{Kraus} {et~al.}(2011){Kraus}, {Senft}, \& {Stewart}}]{2011Icar..214..724K}
{Kraus}, R.~G., {Senft}, L.~E., \& {Stewart}, S.~T. 2011, \icarus, 214, 724

\bibitem[{{Kurosaki} {et~al.}(2023){Kurosaki}, {Hori}, {Ogihara}, \& {Kunitomo}}]{kurosaki+2023b}
{Kurosaki}, K., {Hori}, Y., {Ogihara}, M., \& {Kunitomo}, M. 2023, \apj, 957, 67

\bibitem[{{Kurosaki} \& {Inutsuka}(2023)}]{Kurosaki+2023}
{Kurosaki}, K. \& {Inutsuka}, S.-i. 2023, \apj, 954, 196

\bibitem[{{Landau} \& {Lifshitz}(1959)}]{Landau_fluid}
{Landau}, L.~D. \& {Lifshitz}, E.~M. 1959, in {Fluid mechanics} (Oxford: Pergamon Press)

\bibitem[{{Lange} \& {Ahrens}(1982)}]{Lange_Ahrens_1982Icar}
{Lange}, M.~A. \& {Ahrens}, T.~J. 1982, \icarus, 51, 96

\bibitem[{{Lock} \& {Stewart}(2024)}]{lock+2024}
{Lock}, S.~J. \& {Stewart}, S.~T. 2024, \psj, 5, 28

\bibitem[{{Lupu} {et~al.}(2014){Lupu}, {Zahnle}, {Marley}, {Schaefer}, {Fegley}, {Morley}, {Cahoy}, {Freedman}, \& {Fortney}}]{Lupu+14}
{Lupu}, R.~E., {Zahnle}, K., {Marley}, M.~S., {et~al.} 2014, \apj, 784, 27

\bibitem[{{Manske} {et~al.}(2022){Manske}, {W{\"u}nnemann}, \& {Kurosawa}}]{Manske+2022}
{Manske}, L., {W{\"u}nnemann}, K., \& {Kurosawa}, K. 2022, Journal of Geophysical Research (Planets), 127, e2022JE007426

\bibitem[{{Melosh}(1989)}]{1989icgp.book.....M}
{Melosh}, H.~J. 1989, {Impact cratering : a geologic process} (Oxford: Oxford University Press)

\bibitem[{{Melosh}(2007)}]{2007MAPS...42.2079M}
{Melosh}, H.~J. 2007, M\&PS, 42, 2079

\bibitem[{{Melosh} \& {Vickery}(1989)}]{Melosh_Vickery_1989Natur}
{Melosh}, H.~J. \& {Vickery}, A.~M. 1989, \nat, 338, 487

\bibitem[{{Mizutani} {et~al.}(1990){Mizutani}, {Takagi}, \& {Kawakami}}]{Mizutani_takagi_1990}
{Mizutani}, H., {Takagi}, Y., \& {Kawakami}, S.-I. 1990, \icarus, 87, 307

\bibitem[{{Monteux} \& {Arkani-Hamed}(2016)}]{Monteux_Arkani_2016}
{Monteux}, J. \& {Arkani-Hamed}, J. 2016, \icarus, 264, 246

\bibitem[{{Monteux} \& {Arkani-Hamed}(2019)}]{Monteux+2019}
{Monteux}, J. \& {Arkani-Hamed}, J. 2019, \icarus, 331, 238

\bibitem[{Pierazzo {et~al.}(2005)Pierazzo, Artemieva, \& Ivanov}]{2005..basalt..pierazzo}
Pierazzo, E., Artemieva, N., \& Ivanov, B. 2005, in {Large Meteorite Impacts III} (Geological Society of America)

\bibitem[{{Pierazzo} \& {Melosh}(2000)}]{2000Icar..145..252P}
{Pierazzo}, E. \& {Melosh}, H.~J. 2000, \icarus, 145, 252

\bibitem[{{Pierazzo} {et~al.}(1995){Pierazzo}, {Vickery}, \& {Melosh}}]{Pierazzo_Vickery_1995LPI}
{Pierazzo}, E., {Vickery}, A.~M., \& {Melosh}, H.~J. 1995, in Lunar and Planetary Science Conference, Vol.~26, Lunar and Planetary Science Conference, 1119

\bibitem[{{Pierazzo} {et~al.}(1997){Pierazzo}, {Vickery}, \& {Melosh}}]{1997Icar..127..408P}
{Pierazzo}, E., {Vickery}, A.~M., \& {Melosh}, H.~J. 1997, \icarus, 127, 408

\bibitem[{{Schlichting} {et~al.}(2015){Schlichting}, {Sari}, \& {Yalinewich}}]{schlichting+2015}
{Schlichting}, H.~E., {Sari}, R., \& {Yalinewich}, A. 2015, \icarus, 247, 81

\bibitem[{{Shuvalov} {et~al.}(2014){Shuvalov}, {K{\"u}hrt}, {de Niem}, \& {W{\"u}nnemann}}]{Shuvalov+2014}
{Shuvalov}, V., {K{\"u}hrt}, E., {de Niem}, D., \& {W{\"u}nnemann}, K. 2014, \planss, 98, 120

\bibitem[{{Stewart} {et~al.}(2014){Stewart}, {Lock}, \& {Mukhopadhyay}}]{Stewart+2014}
{Stewart}, S.~T., {Lock}, S.~J., \& {Mukhopadhyay}, S. 2014, in 45th Annual Lunar and Planetary Science Conference, Lunar and Planetary Science Conference, 2869

\bibitem[{{Stewart} {et~al.}(2008){Stewart}, {Seifter}, \& {Obst}}]{Stewart_Seifter_2008GeoRL}
{Stewart}, S.~T., {Seifter}, A., \& {Obst}, A.~W. 2008, \grl, 35, L23203

\bibitem[{{Suetsugu} {et~al.}(2018){Suetsugu}, {Tanaka}, {Kobayashi}, \& {Genda}}]{suetsugu18}
{Suetsugu}, R., {Tanaka}, H., {Kobayashi}, H., \& {Genda}, H. 2018, \icarus, 314, 121

\bibitem[{{Sugita} \& {Schultz}(2002)}]{sugita+2002}
{Sugita}, S. \& {Schultz}, P.~H. 2002, \icarus, 155, 265

\bibitem[{{Thompson} \& {Lauson}(1972)}]{1972snl..rept..714T}
{Thompson}, S.~L. \& {Lauson}, H.~S. 1972, {Improvements in the Chart D Radiation-Hydrodynamic Code. III: Revised Analytic Equations of State}, Albuquerque, New Mexico: Sandia National Laboratory. 1972. Technical Report SC-RR-71-0714

\bibitem[{{Thompson} {et~al.}(2019){Thompson}, {Lauson}, {Melosh}, {Collins}, \& {Stewart}}]{thompson+2019}
{Thompson}, S.~L., {Lauson}, H.~S., {Melosh}, H.~J., {Collins}, G.~S., \& {Stewart}, S.~T. 2019, {M-ANEOS}

\bibitem[{{Tonks} \& {Melosh}(1993)}]{Tonks_melosh_1993}
{Tonks}, W.~B. \& {Melosh}, H.~J. 1993, J.G.R., 98, 5319

\bibitem[{{Turtle} \& {Pierazzo}(2001)}]{2001Sci...294.1326T}
{Turtle}, E.~P. \& {Pierazzo}, E. 2001, Science, 294, 1326

\bibitem[{{W{\"u}nnemann} {et~al.}(2006){W{\"u}nnemann}, {Collins}, \& {Melosh}}]{2006Icar..180..514W}
{W{\"u}nnemann}, K., {Collins}, G.~S., \& {Melosh}, H.~J. 2006, \icarus, 180, 514

\end{thebibliography}

\begin{appendix}
\section{Derivation of $\xi_{\rm s}$}
\label{appendix}
The assumption of a strong explosion with spherical symmetry allows to write 
the integrand in Eq.~(\ref{eq:ene_conserv_dimen}) with 
dimensionless functions of $\xi$ as 
\begin{eqnarray}
    \up(R_{\rm c},t)&=&\frac{2R_{\rm c}}{5t}V(\xi),\label{eq:nondimen_v}\\
    \rho(R_{\rm c},t)&=&\rho_0 G(\xi),\label{eq:nondimen_rho}\\
 E(R_{\rm c},t)&=&\frac{P(R_{\rm c},t)}{2({\cal S}-1)}\\
 &=& \frac{\rho_0}{2({\cal S}-1)} \left(\frac{2R_{\rm c}}{5t} \right)^2 Z(\xi).\label{eq:nondimen_p}
\end{eqnarray}

The functions $V$, $G$, and $Z$ are 
given by solving fluid equations with boundary conditions at $\xi=\xi_{\rm s}$ \citep{Landau_fluid}. 
\begin{eqnarray}\label{eq:exact_nondim}
    \left[\frac{2(3\s-2)V-5}{(\s-4)/\s} \right]^a \left[\frac{(2\s-1)V-1}{(S-1)/\s} \right]^b\nonumber \\ 
    \times(\s V)^{-2/5}=\chi,
\end{eqnarray}
\begin{eqnarray}
    G&=& \frac{\s}{\s-1}\left[\frac{2(3\s-2)V-5}{(\s-4)/\s} \right]^{a^\prime}\nonumber\label{eq:func_g}\\
    &\times&\left[\frac{(2\s-1)V-1}{(S-1)/\s} \right]^{b^\prime}\left[\frac{V-1}{(1-\s)/\s}\right]^{c^\prime},
\end{eqnarray}
\begin{eqnarray}
    Z = (1-\s)V^2\frac{V-1}{(2\s-1)V-1},\label{eq:func_z}
\end{eqnarray}
where $\chi = \xi / \xi_{\rm s}$ and 
\begin{eqnarray}
    a&=&\frac{26\s^2 -33\s+16}{5(3\s-2)(4\s-1)},\\
    b&=&\frac{2(\s-1)}{4\s-1},\\
    a^\prime&=&\frac{26\s^2 -33\s+16}{(3\s-2)(4\s-1)(2\s-3)},\\
    b^\prime&=&3/(4\s-1),\\
    c^\prime&=&2/(2\s-3).\label{eq:exponent_nondim}
\end{eqnarray}

Substituting Eqs.~(\ref{eq:nondimen_v})--(\ref{eq:nondimen_p}) into Eq.~(\ref{eq:ene_conserv_dimen}), we have Eq.~(\ref{eq:ene_conserv_nondim}). 
For the integration in Eq.~(\ref{eq:ene_conserv_nondim}), $V, G$, and $Z$ are obtained as functions of $\chi$, using Eqs.~(\ref{eq:func_g})--(\ref{eq:func_z}).

\end{appendix}

\end{document}